\documentclass[journal]{IEEEtran}

\def\NoNumber#1{{\def\alglinenumber##1{}\State #1}\addtocounter{ALG@line}{-1}}

\ifCLASSINFOpdf

\else

\fi

\newtheorem{theorem}{Theorem}

\usepackage{tikz}

\usepackage{subcaption} 
\usepackage{amsmath}
\usepackage{amsfonts}
\usepackage{amssymb}
\usepackage[bookmarks=false]{hyperref}
\usepackage{orcidlink}
\usepackage{multirow}
\usepackage{graphicx}
\usepackage{textcomp}
\usepackage{xcolor}
\graphicspath{ {./images/} }
\usepackage{booktabs}
\usepackage{lscape}
\usepackage{adjustbox}
\usepackage{algorithmicx}
\usepackage{algorithm}

\newcommand*{\Scale}[2][4]{\scalebox{#1}{$#2$}}%

\newcommand\copyrighttext{%
  \footnotesize This paper has been accepted for publication in a future issue of \href{https://ieeexplore.ieee.org/xpl/aboutJournal.jsp?punumber=49}{IEEE Journal on Selected Areas in Communications (JSAC)}, but has not been fully edited. Content may change prior to final publication. \textcopyright 2023 IEEE. Personal use of this material is permitted. Permission from IEEE must be obtained for all other uses, in any current or future media, including reprinting/republishing this material for advertising or promotional purposes, creating new collective works, for resale or redistribution to servers or lists, or reuse of any copyrighted component of this work in other works.}
\newcommand\copyrightnotice{%
\begin{tikzpicture}[remember picture,overlay]
\node[anchor=north,yshift=-5pt] at (current page.north) {\fbox{\parbox{\dimexpr\textwidth-\fboxsep-\fboxrule\relax}{\copyrighttext}}};
\end{tikzpicture}%
}

\begin{document}

\title{Safe and Accelerated Deep Reinforcement Learning-based O-RAN Slicing: A Hybrid \\ Transfer Learning Approach}

\author{Ahmad M. Nagib\orcidlink{0000-0002-9193-9755}, \textit{Graduate Student Member, IEEE}, Hatem Abou-zeid, \textit{Member, IEEE}, \newline and Hossam S. Hassanein\orcidlink{0000-0003-0260-8979}, \textit{Fellow, IEEE}
        
\thanks{This research was supported by the Natural Sciences and Engineering Research Council of Canada (NSERC) under Grant RGPIN-2019-05667 and Grant RGPIN-2021-04050. (Corresponding author: Ahmad M. Nagib.) 

Ahmad M. Nagib is with the School of Computing, Queen’s University, Kingston, ON K7L 2N8, Canada, and also with the Faculty of Computers and Artificial Intelligence, Cairo University, Giza 12613, Egypt (e-mail: ahmad@cs.queensu.ca). 

Hatem Abou-Zeid is with the Department of Electrical and Software Engineering, University of Calgary, Calgary, AB T2N 1N4, Canada (e-mail: hatem.abouzeid@ucalgary.ca).

Hossam S. Hassanein is with the School of Computing, Queen’s University, Kingston, ON K7L 2N8, Canada (e-mail: hossam@cs.queensu.ca).} 
        }

\maketitle
\copyrightnotice

\pagenumbering{gobble}

\begin{abstract}

The open radio access network (O-RAN) architecture supports intelligent network control algorithms as one of its core capabilities. Data-driven applications incorporate such algorithms to optimize radio access network (RAN) functions via RAN intelligent controllers (RICs). Deep reinforcement learning (DRL) algorithms are among the main approaches adopted in the O-RAN literature to solve dynamic radio resource management problems. However, despite the benefits introduced by the O-RAN RICs, the practical adoption of DRL algorithms in real network deployments falls behind. This is primarily due to the slow convergence and unstable performance exhibited by DRL agents upon deployment and when encountering previously unseen network conditions. In this paper, we address these challenges by proposing transfer learning (TL) as a core component of the training and deployment workflows for the DRL-based closed-loop control of O-RAN functionalities. To this end, we propose and design a hybrid TL-aided approach that leverages the advantages of both policy reuse and distillation TL methods to provide \textit{safe and accelerated} convergence in DRL-based O-RAN slicing. We conduct a thorough experiment that accommodates multiple services, including real VR gaming traffic to reflect practical scenarios of O-RAN slicing. We also propose and implement policy reuse and distillation-aided DRL and non-TL-aided DRL as three separate baselines. The proposed hybrid approach shows at least: 7.7\% and 20.7\% improvements in the average initial reward value and the percentage of converged scenarios, and a 64.6\% decrease in reward variance while maintaining fast convergence and enhancing the generalizability compared with the baselines.

\end{abstract}

\begin{IEEEkeywords}
Deep Reinforcement Learning (DRL), Transfer Learning (TL), Trustworthy DRL, Safe and Accelerated DRL, O-RAN Slicing, 6G
\end{IEEEkeywords}

\section{Introduction}

\IEEEPARstart{T}{he} open radio access network (O-RAN) architecture \cite{o-ran-specification} was proposed by the O-RAN alliance to support the evolution of next-generation networks (NGNs) \cite{9579445}. Virtualization, openness, and intelligence are inherent properties of such an architecture. The O-RAN architecture provides open interfaces for flexible network management and automation \cite{9839628}. These standardized interfaces will enable mobile network operators (MNOs) to flexibly and intelligently control various radio resource management (RRM) functionalities in a closed-loop fashion. The flexibility provided by O-RAN is essential as more customizable radio access network (RAN) products are needed to adapt to the various network scenarios and the new services offered. This also allows MNOs to dynamically change the network configurations to reflect their priorities and objectives at a given time. The O-RAN paradigm is therefore expected to bring gains in many cellular network applications, especially those that necessitate dynamic control based on the network conditions and service requirements such as network slicing \cite{9839628}. In NGNs, the optimization domains and network requirements are expected to become larger and tighter respectively. This will make solving dynamic RRM problems even more complex \cite{8466370}.

Next-generation cellular networks have key performance indicators (KPIs) and other measurements used to quantify the performance of the various services. Examples are throughput, latency, quality of experience (QoE), quality of service (QoS), and radio channel conditions. This is seamlessly compatible with the deep reinforcement learning (DRL) feedback loop of observing the system state, taking action, and receiving rewards accordingly. DRL agents can adapt to the dynamic O-RAN environment and make quick decisions based on the available knowledge in an open-control fashion \cite{9372298}. Hence, DRL algorithms are among the most promising methods to design O-RAN-compliant data-driven applications hosted by the near-real-time (near-RT) RAN intelligent controllers (RICs) \cite{9812489}. Such applications are called xApps. This allows an MNO to intelligently adjust network settings to achieve an optimal RRM configuration for a given network condition.

Despite the potential benefits introduced by the O-RAN RICs, the practical adoption of DRL algorithms in real network deployments falls behind. The main reasons for this are the slow convergence and unstable performance that the DRL agents undergo \cite{9430561}. This is particularly evident when DRL-based xApps are newly deployed and when experiencing extreme situations or significant changes in the network conditions \cite{9903386}. Slow convergence relates to the considerable number of time steps the DRL agent takes to find or recover optimal configurations for a given RRM functionality. Unstable performance relates to sudden drops in the O-RAN system's performance. A certain performance level must be maintained by O-RAN systems to guarantee users’ QoE and the overall system’s QoS. Hence, the instabilities due to DRL exploration affect these two measurements negatively.

The training of DRL agents should be done offline initially according to the O-RAN recommendations \cite{10071941}. This ensures that the trained models do not affect the performance and stability of the network. Nevertheless, the offline simulation environments are usually inaccurate and do not reflect all the situations that could be experienced in practical deployment environments. This applies even if real network data was logged and used to simulate such environments offline \cite{9061001}. This is not the case in other applications such as training a DRL agent to play a computer game. Unlike O-RAN-based NGNs, the game training environment will still match the deployment environment. However, whenever an xApp is newly deployed in O-RAN's near-RT RIC, some online learning is still required by the incorporated DRL agent to adapt to the live network environment \cite{9931127}. 

Moreover, learning is needed whenever the agent experiences extreme cases or when the network context changes significantly \cite{9229155}. In both situations, some exploration may be required, while the DRL agent recovers, to avoid affecting the performance of the available services substantially. Convergence needs to be quick and stable so that the end user's QoE is not affected. Nonetheless, it may take thousands of learning steps to recover, given the stochasticity of NGNs O-RAN systems and the exploratory nature of the DRL-based xApps. This is of great significance in live network deployments. NGNs can only tolerate a few iterations of stable exploration while optimizing near-RT O-RAN functionalities \cite{9430561}.

The approaches used to tackle such challenges are known as \textit{safe and accelerated} DRL techniques as defined in \cite{9903386} and \cite{10.5555/2789272.2886795}. Such techniques attempt to reduce service level agreements (SLAs) violations and to avoid any system performance instabilities. They also aim to shorten the exploration and recovery duration of DRL-based xApps in O-RAN. These approaches can help pave the way for adopting trustworthy DRL to optimize dynamic RRM functionalities in O-RAN. Transfer learning (TL) is among the main methods used to address the DRL-related practical challenges mentioned earlier \cite{9789336, 9999297}. TL can be used to guide a newly deployed DRL-based xApp while learning the optimal policy in network conditions it has not experienced before. A policy learned by a previously trained DRL agent can be used as a guide in such a case. This can be done in several ways as demonstrated in this study. 

In this paper, we address the challenge of slow and unstable DRL convergence in the context of O-RAN slicing. To the best of our knowledge, this is the first work to propose TL as a core component for \textit{safe and accelerated} DRL-based xApps in O-RAN, and more specifically in the closed-loop control of O-RAN slicing. Our contributions can be summarized as follows:
\begin{itemize}
    \item We propose to incorporate TL as a core component of DRL-based control of network functionalities in the O-RAN architecture. We propose training and deployment workflows in the non-real-time (non-RT) and near-RT RICs respectively. TL tackles the challenges of slow and unstable DRL convergence by reusing knowledge from pre-trained expert policies. The proposed flows aim at enhancing the convergence and generalizability of O-RAN DRL-based xApps. They accommodate the difference between offline training and live deployment environments. They also accommodate significant changes in the network conditions.

    \item We propose a hybrid TL-aided DRL approach for \textit{safe and accelerated} convergence of DRL-based O-RAN slicing xApps. The proposed approach combines policy reuse and distillation TL methods to guide the DRL agent's convergence and strike a balance between deterministic and directed exploratory actions. We also propose policy reuse and distillation as two separate TL-aided DRL baselines in addition to the non-TL-aided DRL baseline.
    
    \item We conduct a thorough study on intelligent O-RAN slicing to demonstrate the gains of the proposed hybrid TL-aided DRL approach. We analyze the reward convergence behavior of the proposed approach and baselines. We then evaluate their safety and acceleration aspects. We finally investigate the effect of the introduced parameter, $\gamma$, which controls the TL method to be used, on the performance of the proposed hybrid approach. Our approach shows at least: a 7.7\% and 20.7\% improvements in the average initial reward value and the percentage of converged scenarios, and a 64.6\% decrease in reward variance while maintaining fast convergence and enhancing the generalizability compared with the baselines.
    
    \item Our experiments support multiple services, including real VR gaming traffic to reflect immersive scenarios of O-RAN slicing in NGNs. We develop and publicly share the implementation of our OpenAI Gym-compatible DRL environment, and the proposed approach and baselines to facilitate research on trustworthy DRL in O-RAN.
    
\end{itemize}

The rest of the paper is structured as follows: Section \ref{relatedwork} discusses the related work. Section \ref{system} details the system model. The proposed O-RAN workflows, hybrid approach, and baselines are described in Section \ref{flows-section}. Section \ref{sec:simulations} includes the experimental setup and an analysis of the results. Finally, we conclude our work and present potential future directions in Section \ref{conclusions}.

\section{Related Work}
\label{relatedwork}

The work in \cite{ferrus2020machine} is amongst the earliest research to consider slicing from an O-RAN perspective. The authors promote using DRL as one of several machine learning (ML) schemes to optimize O-RAN slicing. They highlight that DRL provides faster convergence when compared with tabular reinforcement learning (RL) given large state and action spaces. The authors of \cite{9999295} and \cite{10001658} employ the concept of federated reinforcement learning (FRL) in the context of O-RAN slicing. In \cite{9999295}, the authors suggest that the global model built using FRL learns to generalize at a slow rate. However, it achieves relatively more robust training by leveraging the shared experience. In \cite{10001658}, the authors employ the knowledge gained in one network application and share it to solve a slightly different problem in another application. This is done by coordinating power control and radio resource allocation xApps for network slicing in O-RAN. A joint global model is created and then disassembled into local Q-tables for the different xApps to follow when deciding the actions to take. The authors indicate that FRL can enable faster convergence but with relatively lower rewards. Nevertheless, a global generic model can still be prone to instabilities and require some exploration when transferred to make decisions in a target local context.

The concept of \textit{safe and accelerated} DRL is partially addressed by a few research studies in the context of O-RAN. In \cite{9903386}, we discussed the need for and categorized the approaches to \textit{safe and accelerated} DRL in NGNs. We also examined the viability of TL variants to accelerate DRL-based RAN slicing using a basic case study. In \cite{9771605}, the problem of resource allocation in O-RAN slicing is addressed. The authors mainly rely on the inherent properties of the DRL algorithms when choosing one to employ. For instance, they mention that the implemented actor-critic (AC)-based solution can produce a faster convergence compared to the proximal policy optimization (PPO)-based one as the off-policy model has relatively improved sample efficiency. However, they show that the AC-based solution has a failed exploration at the beginning and a lower reward value in general. Moreover, both DRL-based solutions can take around 20 thousand learning steps to converge with no guarantee of fast and stable performance if deployed in the near-RT RIC of O-RAN. Hence, robust performance cannot be insured solely based on the inherent properties of the DRL algorithm chosen. 

In \cite{9814869}, O-RAN slicing is one of the 3 developed exemplary xApps. The authors address some convergence-related issues by choosing PPO as it proved to be reliable and efficient in literature. They also propose pre-processing the observations using autoencoders before feeding them into the DRL agent. This reduces the dimensionality of the observations yet retains a good representation of the system state. Moreover, in \cite{10078092}, the authors show that the proposed deep Q-network (DQN)-based algorithm experiences relatively slow convergence on the communication level of slicing compared with the computational level. They indicate that this is primarily due to the mobility of end devices. Consequently, the channel gains between the base stations (BSs) and the end devices change frequently, delaying the convergence. More relevantly, authors of \cite{10008614} attempt to accelerate the convergence by proposing an evolutionary-based DRL approach in the context of O-RAN slicing. However, it appears that training and deployment flows are not isolated. Meaning that the DRL training is carried out in the near-RT RIC. Online training is not recommended by the O-RAN alliance as it comes with risks such as the requirement for exploration \cite{9931127}. Furthermore, the costs of online training are not demonstrated.

The authors of \cite{10.1145/3551660.3560908} propose a centralized DRL-based approach for dynamic RAN slicing. They run simulations using different combinations of parameters to find those leading to short convergence times. This is mainly done for the offline training process and still falls under the ``acceleration via design choices" category defined in \cite{9903386}. They also propose a non-DRL-based crowding game approach that experiences relatively faster convergence in the offline training phase. Nonetheless, if deployed in the near-RT RIC, additional time is required for the crowding game approach to compute the cost of each strategy and make a decision as it is based on the users' KPIs. Finally, in \cite{9931127}, parallelization is suggested as one of the design approaches for reducing the convergence time. This is mainly done by utilizing several environments in parallel. This can reduce the real clock time of DRL agents' convergence during offline training. However, this is not supported during the deployment phase in real networks. 

Different from the aforementioned reviewed work, in this paper, we propose isolated systematic training-deployment workflows that consider the O-RAN recommendations. Such flows incorporate expert policies pre-trained on the same problem, and fine-tuned using live network data, to guide the DRL convergence. The flows are algorithm-agnostic and should work with any DRL algorithms or settings configured by the MNO. Furthermore, the proposed policy transfer-aided DRL approach and baselines are primarily concerned with the practical live deployment in the O-RAN's near-RT RIC. Finally, the proposed hybrid TL-aided DRL approach ensures that the maximum level of rewards is reached safely and quickly. We demonstrate that using real network VR gaming traffic to reflect an important immersive and latency-intolerant scenario in NGNs \cite{9685808}.

\begin{figure*}
\centering
\includegraphics[width=0.7\linewidth]{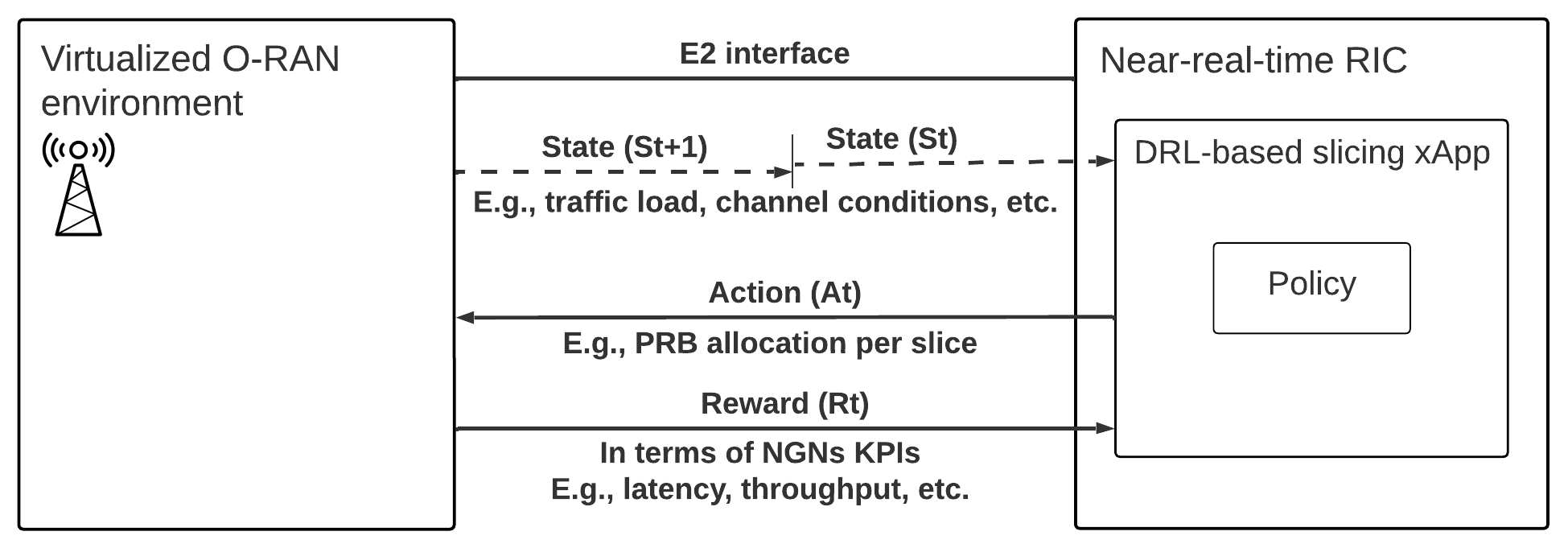}
\caption{Block diagram of the DRL-based slicing xApp interaction with the O-RAN environment.}
\label{fig:drl-normal-flow}

\end{figure*}

\section{Deep Reinforcement Learning-based O-RAN Slicing}
\label{system}

\subsection{O-RAN Intelligent Controllers}

O-RAN will give MNOs more control over the network. For instance, the O-RAN-based NGNs will include generic modules and interfaces for data collection, distribution, and processing \cite{9839628}. This will enable data-driven closed-loop control using ML as a core component of the network operation \cite{9627832}. Such data-driven applications can be deployed on two levels, namely, near-RT RIC, and non-RT RIC. They are called xApps when deployed on the near-RT RIC and rApps on the non-RT RIC \cite{10071941}. xApps interact with the RAN nodes via the E2 interface. The E2 interface includes the service model (SM) component that helps xApps fulfill the various RRM functionalities. This can be accomplished through a standardized interaction between xApps and the virtualized O-RAN infrastructure as shown in Fig. \ref{fig:drl-normal-flow}.

NGNs will be heterogeneous on multiple levels. This includes radio access technologies (RATs), communication paradigms, and cell and user equipment (UE) types. O-RAN slicing is one of the paradigms that enable extra flexibility for MNOs. It allows for supporting a wide range of use cases and deployment scenarios simultaneously \cite{7891795}. This happens while sharing the same infrastructure among various services in a way that fulfills their different requirements. Hence, MNOs are required to optimize a myriad of network functionalities that operate at different timescales and have different goals \cite{8466370}. For instance, admission control, packet scheduling, and handover management are examples of network functionalities that require the efficient utilization of scarce radio resources \cite{9372298}. This process is called RRM \cite{9430561} and a common approach to managing such limited radio resources is to build a DRL-based O-RAN xApp. As illustrated in Fig. \ref{fig:drl-normal-flow}, a DRL agent observes the state of the O-RAN environment and takes action with the objective of maximizing its rewards. This is measured in terms of network KPIs relevant to the functionality at hand and this process is carried out in an open-control fashion.

O-RAN standardized open interfaces enable MNOs to collect live data from the RAN to optimize network performance. This vital paradigm will be more feasible on a larger scale in future 6G networks. Such O-RAN interfaces allow the consultation of DRL agents to select the best actions to optimize a network function given a network condition as highlighted in Fig. \ref{fig:drl-normal-flow}.

\subsection{System Model}
\label{model}

In this paper, we focus on the downlink O-RAN inter-slice resource allocation problem. This belongs to the radio access level of network slicing as defined in \cite{7891795}. The objective is to allocate the available physical resource blocks (PRBs) to the admitted slices while fulfilling their SLAs. A list of some notations used in this paper is provided in Table \ref{tab:symbols}. The inter-slice radio resource allocation problem can be formulated as follows \cite{9430561, 9524965}:

Any given BS supports multiple services that are reflected by a set of slices, $\mathcal{S}=\{1,2, \ldots, S\}$, that share the available bandwidth, $B$. We consider a set of UEs, $\mathcal{U}=\{1,2, \ldots, U\}$, connected to a BS. Each UE, $u$, can request one type of service at a time for downlink transmission. A slice, $s$, has a set of requests, $\mathcal{R}_s=\left\{1,2, \ldots, R_s\right\}$, where $R_s$ is the number of requests made by users belonging to a slice $s$. The total demand, $D_s$, of such users can be represented as follows:
\begin{equation}
D_s=\sum_{r_s \in R_s} d_{r_s},
\end{equation}

where $d_{r_s}$ is the demand of a request, $r_s$, made by a user belonging to slice $s$. Moreover, any given slice, $s$, contributes to the overall BS's traffic as follows:
\begin{equation}
\label{contribution}
\kappa_s=\frac{D_s}{\sum_{i=1}^{\|S\|} D_i}
\end{equation}

The allocation of PRBs among the available slices, $S$, needs to be optimized. This can be described by the vector, ${a \in \rm I\!R ^S}$. At the beginning of any slicing window, an O-RAN slicing xApp decides to choose a specific slicing PRB allocation configuration, ${a}$, out of the $A$ possible configurations, where $\mathcal{A}=\{1,2, \ldots, A\}$. Based on such a decision, the system performance is affected. For the purpose of this paper, the system performance is represented in terms of the latency of the admitted slices. This mainly depends on a queue maintained at the BS.

\begin{table}

\centering
  \caption{List of Notations}
  \label{tab:symbols}
\begin{tabular}{|p{0.32in}|p{2.8in}|}
    \hline
 \textbf{Symbol}&\textbf{Description}\\
    \hline
    \textbf{\hfil$S$}& Number of available slices\\
    \hline
    \textbf{\hfil$B$}& Available bandwidth shared among slices \\ 
    \hline
    \textbf{\hfil$U$}& Number of available UEs\\
    \hline
    \textbf{\hfil$R_s$}&  Number of requests made by users belonging to a slice $s$\\ 
    \hline
    \textbf{\hfil$D_s$}& Total demand of users belonging to a slice $s$\\ 
    \hline
    \textbf{\hfil$d_{r_s}$}& Demand of a request $r_s$ made by a user of slice $s$ \\ 
    \hline
    \textbf{\hfil$\kappa_s$}& Contribution of slice $s$ to overall BS's traffic\\
    \hline
    \textbf{\hfil$\mathcal{A}$}& Set of available slicing PRB allocation configurations\\ 
    \hline
    \textbf{\hfil$a$}& A given slicing PRB allocation configuration\\
    \hline
    \textbf{\hfil$b_{s}$}& Bandwidth allocated to slice $s$\\
    \hline
    \textbf{\hfil$R$}& Reward function\\
    \hline
    \textbf{\hfil$w_{s}$}& Priority of fulfilling the latency requirement of slice $s$\\
    \hline
    \textbf{\hfil$\Omega$}& Slicing window size\\
    \hline
    \textbf{\hfil$l_{s}$}& Average latency in the previous slicing window for slice $s$\\
    \hline
   \textbf{\hfil$\pi$}& RL agent's policy\\ 
    \hline
    \textbf{\hfil$\pi_{\text{E}}$}& Expert policy\\
    \hline 
    \textbf{\hfil$\pi_{\text{L}}$}& Learner policy\\
    \hline
    \textbf{\hfil$\mathcal{P}_{\text{E}}$}& Set of stored expert policies\\
    \hline
    \textbf{\hfil${c}_{1}$}& A sigmoid function parameter to decide the point to start penalizing the agent's actions\\ 
    \hline
    \textbf{\hfil${c}_{2}$}& A sigmoid function parameter to reflect the acceptable latency for each slice\\ 
    \hline
    \textbf{\hfil$\mathcal{M}$}&Source domain\\
    \hline
    \textbf{\hfil$T$}& Knowledge transfer duration\\ 
    \hline
    \textbf{\hfil$\theta$}& Transfer rate which decides whether to follow the transferred knowledge or the learner policy \\ 
    \hline
    \textbf{\hfil$\nu$}& Transfer rate decay \\ 
    \hline
    \textbf{\hfil$\gamma$}& Hybrid approach parameter which decides the policy transfer method to follow\\ 
    \hline
\end{tabular}
\end{table}

\subsection{O-RAN Slicing: Mapping to Deep Reinforcement Learning}
\label{mapping}

A DRL-based xApps's objective is to maximize the long-term reward expectation, that is,

\begin{equation} \label{eq:expect}
\begin{aligned}
\underset{a}{\text{argmax}}\ \mathbb{E}\{R(\boldsymbol{a}, \boldsymbol{\kappa})\},
\end{aligned}
\end{equation}

where $\mathbb{E}(\cdot)$ represents the expectation of the argument. This allows us to learn a policy, $\pi$, that takes a state, ${\kappa\in \rm\! \mathcal{K}}$, as input, and outputs an action, ${a = \pi(\kappa) \in  \! \mathcal{A}}$. The main challenge in solving (\ref{eq:expect}) is the varying demand over time. To find the optimal solution, an exhaustive search can be performed, considering all possible allocations at the start of each slicing window and recording the resulting system performance. However, this approach is both computationally expensive and practically infeasible. Therefore, DRL provides a viable alternative for solving the problem. We describe our DRL design in the following subsections.

\subsubsection{State Representation}

As seen in Fig. \ref{fig:drl-normal-flow}, the slicing xApp deployed in the near-RT RIC begins with observing the system state. We represent the state of the O-RAN system in terms of the slices' contribution to the overall BS's traffic within the preceding slicing window, ${\Omega}_{t-1}$. This can be reflected by a vector of size $S$ as follows:


\begin{equation}
\label{eq:state-represenation}
\kappa = (\kappa_{1}, . . . , \kappa_s, . . . ,\kappa_{S}) 
\end{equation}

\subsubsection{Action Space}

Based on the observed state, the xApp takes an action at the beginning of each slicing window. It selects the PRB allocation configuration per slice. We represent it as the percentage of bandwidth allocated to each slice as follows:
\begin{equation}
\label{eq:action}
a = (b_{1}, . . . , b_{s}, . . . ,b_{S}), 
\text { subject to } b_{1} + . . . + b_{S} = B 
\end{equation}

\subsubsection{Reward Function Design}

After taking the action, the DRL-based xApp receives reward feedback in terms of network KPIs calculated at the end of every slicing window. In this paper, we define rewards as a function of latency because we prioritize the delay-intolerant VR gaming service and for better results' interpretability. 

Safe RL can be defined as the process of learning policies that maximize the expectation of the return to ensure reasonable system performance or respect safety constraints \cite{10.5555/2789272.2886795}. This can be during the learning or deployment processes. Hence, in this paper, safety can be described as having a reasonable latency performance during the deployment of DRL-based O-RAN slicing xApps. Safe RL can reduce or prevent undesirable situations through 1) transforming the optimization criterion, or 2) modifying the exploration process of the RL agent \cite{10.5555/2789272.2886795}. In this paper, we design a risk-sensitive reward function. In risk-sensitive approaches, the optimization criterion is changed to include a parameter that allows the sensitivity to the risk to be controlled \cite{10.5555/2789272.2886795}. Thus, we employ a sigmoid-based \cite{Leibovich} reward function that includes parameters to reflect the acceptable latency for each slice. This enables penalizing the xApp for undesirable actions that get the system close to violating the defined latency requirements of each slice.

The reward function defined in this study reflects a weighted sum of an inverse form of latency. It allows more control over the effect of getting closer to the minimum acceptable level of each slice's SLAs as follows:


\begin{equation}
\label{eqn:rewardequation}
\begin{aligned}
 R = \sum_{s=1}^{\|S\|} w_{s}\ *\ \frac{\mathrm{1} }{\mathrm{1} + e^{\ c1_{s}\ *\ (\ {l}_{s}\ -\ c2_{s}\ )} }
\end{aligned}
\end{equation}

\begin{figure}
\centering
\includegraphics[width=1\linewidth]{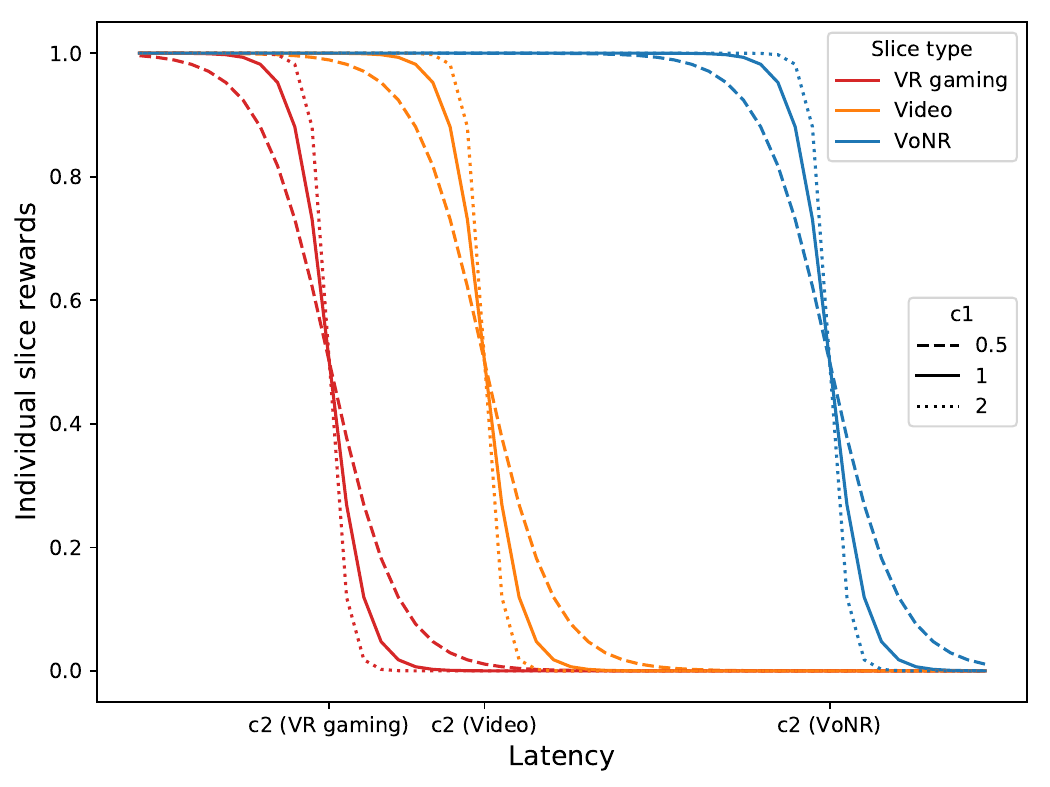}
\caption{An example of the reward function: ${c}_{1}$ decides the point to start penalizing the agent's actions; and ${c}_{2}$ reflects the acceptable latency for each slice.}
\label{fig:sigmoid_function}
\end{figure}

Since we focus on the delay requirements of the different services, we use latency as a variable. The weight, $w_{s}$, reflects the priority of fulfilling the latency requirement of slice $s$, and $l_{s}$ is the average latency experienced within slice $s$ during the previous slicing window. The function's effect can be adjusted by configuring two parameters, namely $c_{1}$ and $c_{2}$ as seen in Fig. \ref{fig:sigmoid_function}. The parameter $c1$ sets the slope for the sigmoid function, thereby indicating when penalties should start being applied to the agent's actions. On the other hand, $c_{2}$ represents the inflection point. Such a point reflects the minimum acceptable delay performance for each slice according to its respective SLAs. Different constant values of $c_{1}$ and $c_{2}$ are utilized for the slices based on the defined SLAs.

\section{Transfer Learning for Safe and Accelerated DRL-based O-RAN Slicing}
\label{flows-section}

We propose to modify the DRL exploration process to avoid risky situations. We do so by including prior knowledge of the learning task by exploiting expert pre-trained policies to allow for a faster and safer DRL exploration setting \cite{10.5555/2789272.2886795}. We propose to incorporate transfer learning as a core component of the training-deployment workflows of DRL-based xApps in the O-RAN architecture. In this section, we first describe the proposed training and deployment flows in the context of O-RAN. Then, we present the developed baselines and the proposed hybrid TL-aided DRL approach.

\begin{figure*}
	\centering
	\includegraphics [width=0.63\linewidth]{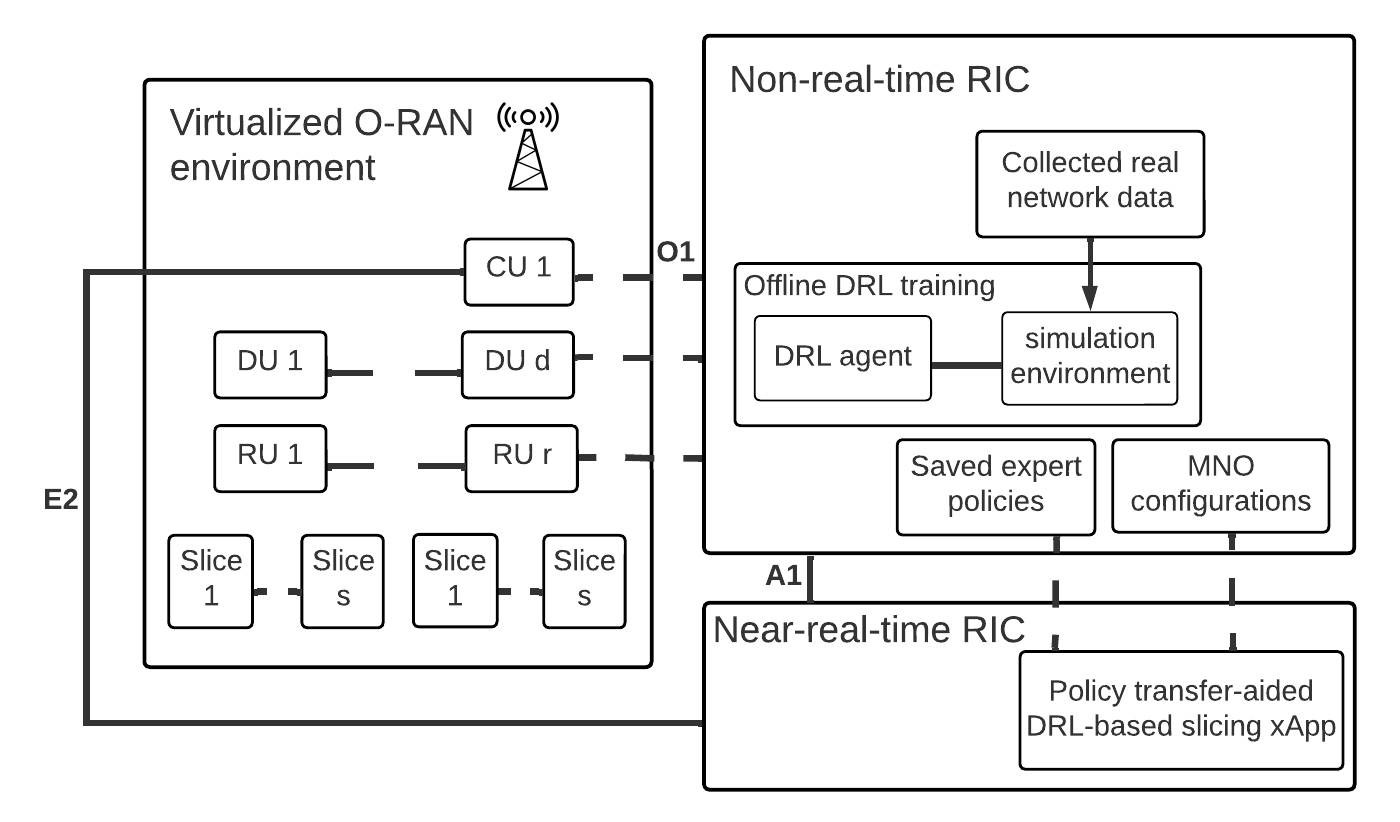}
 \caption{Block diagram of the policy transfer-guided O-RAN system architecture.}
	\label{TL-guided}
\end{figure*}

\subsection{Training and Deployment Flows in Policy Transfer-Aided O-RAN Architecture}
\label{flows}

The DRL training is normally carried out using a simulated offline environment. Hence, when the DRL agent is deployed in a live network, there will be a performance gap leading to an undesired exploration performance \cite{9372298}. This also happens when the context of the network changes significantly. For instance, when the number and type of the available slices change. We propose training and deployment workflows for DRL-based xApps in the O-RAN architecture. These flows address the challenges of slow and unstable DRL convergence in O-RAN-based NGNs. The DRL training and deployment workflows are proposed to be hosted in O-RAN non-RT and near-RT RICs respectively. They aim to enhance the DRL convergence and generalizability in the context of the O-RAN architecture. They also provide the readers with insights on how to overcome key challenges when deploying an O-RAN DRL-based xApp for network slicing, and RRM in general.

\subsubsection{Training Workflow}

The proposed O-RAN training workflow makes use of real network data collected at the non-RT-RIC to train the policy of a learner agent planned for deployment as seen in Fig. \ref{TL-guided}. It does not rely on pure offline simulations or mathematical models in training. Nevertheless, the DRL agent training is still carried out in the non-RT RIC according to O-RAN alliance recommendations as seen in the figure \cite{9814869}. The O1 interface is employed to collect data every $\Omega$ seconds, reflecting the slicing window size. The data collected represents the relevant network measurements during a slicing window. This includes but is not limited to, throughput, delay, number of available slices, types of services supported, and traffic load for each slice. The number of PRBs allocated to each slice should also be logged. The compiled data mainly reflect the system state, action taken, and reward parameters. This allows building offline simulations to train a DRL agent in the non-RT RIC using such data. However, this data does not guarantee that all the state-action pairs are represented. Hence, the training environment still does not reflect all the cases that a DRL agent can experience if deployed in an xApp in the near-RT RIC.

In the training phase, several DRL agents are trained using the collected data to reflect various contexts. The DRL agent's actions are taken based on the system’s state. Rewards are calculated from the collected KPIs that correspond to the logged state-action pairs. This is done until the agents being trained converge. The compiled data should reflect BSs having different contexts and properties. As an initial step, the MNO can store a set of expert policies, $\mathcal{P}_{\text{E}}=\{1,2, \ldots, \pi_{\text{E}}, \ldots, \Pi_{\text{E}}\}$, that result in good convergence performance for the various contexts during the training process. Subsequently, they will be loaded to guide the convergence of other DRL agents via policy transfer. Such a set of policies should also be updated based on the policies' performance after being fine-tuned in a live network setting. 

As proposed in the next sub-section, policy transfer is carried out whenever a new DRL-based xApp is deployed or when the BS context changes. The context can be in the form of the number of slices, types of services supported by the BS, and the MNO's SLA fulfillment priorities at a given time. An approach to choosing the right policy for a given context is another research direction on efficient policy transfer \cite{10075524}.

\subsubsection{Deployment Workflow}

Given a live network context, a policy of a trained DRL agent is deployed as an xApp. Such a DRL-based xApp will still experience some exploration due to the difference between the training and deployment environments \cite{9931127}. Upon the termination of the training phase, the xApp loads the proper policy from the policy directory in the non-RT RIC via the A1 interface as highlighted in Fig. \ref{TL-guided}. The policy is loaded based on the context of the BS to be controlled. Such a policy is used to guide the DRL agent of a newly deployed xApp. This also allows the policies to get fine-tuned using live network data. Once they prove to meet the various slices’ SLAs in certain network contexts, the policy directory should be updated.

In addition to being newly deployed, the xApp may also experience extreme conditions that were not reflected in the training data. The proposed deployment flow reuses existing knowledge from saved expert policies that were proven to provide reasonable performance in certain live network contexts. This accommodates the difference between the training data and the actual live network conditions. This also accommodates significant changes in the network context. The expert policies are used as guidance for the current agent while trying to recover instead of randomly exploring the action space.

Furthermore, an MNO may decide to change some DRL-related configurations. The MNO can do so through the open interfaces supported by O-RAN. For instance, the MNO can reconfigure the slicing window size, state representation, action space, or reward function. As an example, the MNO may decide to modify the weights of the utilized reward function. Such modifications can reflect a change in the MNO's priorities of fulfilling the SLAs of the different network slices. In both situations of extreme conditions and MNO reconfigurations, a new policy can be loaded from the policy directory to match the latest context of the BSs of interest. One or more policies can be loaded at once depending on the policy transfer configurations set by the MNO. Such configurations are inputted to the O-RAN slicing xApp via the A1 interface as seen in Fig. \ref{TL-guided}. Upon deployment, the recommended action is decided based on the system state, and the previously mentioned MNO configurations. This should be done within the time range of the near-RT RIC \cite{10.1145/3551660.3560908}. The xApp then executes the action taken via the E2 interface to allocate resources among the available slices for the duration of $\Omega$ seconds. Accordingly, scheduling is carried out per slice based on the scheduling algorithm configured by the MNO. Then, the state of the system per slicing window is captured based on the state representation chosen by the MNO. Finally, the DRL-based xApp's policy is updated depending on the selected DRL algorithm and other settings such as buffer size, and learning rate. 

Loading a new policy to guide the DRL-based xApp can also be triggered by the reward feedback. For instance, if the reward value drops below a pre-defined threshold value for some time, this may indicate that the context has significantly changed. Hence, the used DRL agent’s policy needs an update. This gives an example of how to identify significant changes in the network context. This will consequently lead to incorporating a new expert policy to guide the DRL agent toward convergence. The questions of how to identify significant changes in network conditions, when to load a new policy, and which policy to use for TL-aided DRL slicing are not the focus of this paper. However, we conducted another study to address a subset of these topics \cite{10075524}. The loaded policy can guide the DRL-based xApp in several ways. This should be decided by the MNO as O-RAN supports customizing the network based on the MNO's preferences. We propose and evaluate the performance of three transfer learning-based approaches in guiding the DRL-based xApps in the following sub-sections.

\subsection{Policy Transfer Baseline Approaches}

In this paper, we propose to employ TL to address the challenge of slow convergence and lack of generalizability of DRL-based xApps. TL can also indirectly tackle the instabilities experienced during convergence. By modifying the exploration process, TL-aided DRL can avoid risky situations by receiving guidance based on prior knowledge. The difference between one type of TL and another mainly depends on the form of knowledge to be transferred. Policy transfer is a type of TL where policies of pre-trained DRL agents are transferred from one or more source domains to guide a newly deployed DRL agent's policy. In this subsection, we first propose to employ two variants of policy transfer, namely, policy reuse and distillation as baselines. We then propose a novel policy transfer method which is a hybrid of such two approaches to achieve an improved convergence performance.

\subsubsection*{Preliminaries}
In general, policy transfer can be carried out via directly reusing expert policies to guide a target learner agent. Alternatively, this can be done via distilling previously acquired knowledge. This can be obtained from the target learner's perspective, or from the source expert's perspective \cite{10172347}. In this paper, we focus on policy distillation from the expert perspective. The knowledge transferred in both policy reuse and distillation approaches is the same. Here, the policy of one or more pre-trained DRL agents is used to guide a newly deployed DRL-based xApp. The main difference between the two approaches is how the transferred policies are used to guide the newly deployed agent to take action. Given a source expert policy $\pi_{E}$ that is trained on data from a source domain $\mathcal{M}$. A learner policy $\pi_{L}$ is trained on data from a target domain guided by the knowledge obtained from $\left\{\pi_{E}\right\}$. When more than one source expert policy is used, a more generic case can be described as follows \cite{10172347}: Given a set of source policies $\pi_{E_{1}}, \pi_{E_{2}}, \ldots, \pi_{E_{P}}$ trained on data from a set of source domains $\mathcal{M}_{1}, \mathcal{M}_{2}, \ldots, \mathcal{M}_{K}$. A learner policy $\pi_{L}$ is trained on data from a target domain by making use of knowledge from $\left\{\pi_{E_{i}}\right\}_{i=1}^{P}$.

\subsubsection{Policy Reuse}

The first policy transfer technique that we propose to employ for accelerating the DRL-based slicing xApp is known as policy reuse. This can be done in several ways \cite{10172347}. In this paper, we propose to carry out policy reuse as described in Algorithm \ref{alg:one}. 

\begin{algorithm}
\caption{Proposed Policy Reuse Approach} 
\label{alg:one}
\textbf{Input:} $\pi_{E}$, MNO configurations, $\kappa$\\
\textbf{Parameters:} $\theta$, $T$, $\Omega$, buffer size, $\beta$, transfer rate decay, $\nu$\\
\textbf{Output:} PRB Allocation per slice
\begin{algorithmic}[1]
\State Load the appropriate pre-trained expert policy from the stored policies $\mathcal{P}_{\text{E}}$
\State Initialize the learner action value function with random weights or from a policy pre-trained using a significantly different traffic pattern
\State \textbf{if} $t < T$ \textbf{do}:
        \State \quad Generate a random number $x$, where $0\leq x \leq 1$
        \State \quad \textbf{if} {$x$ $\leq$ $\theta$} \textbf{do}:
            \State \quad  \quad Consult the expert policy
            \State \quad \quad  Choose an action according to Theorem \eqref{eqn:reuseequation}
       \State \quad \textbf{else if}{ $x > \theta$} \textbf{do}:
            \State \quad \quad Choose an action according to the learner policy
    \State \quad \textbf{end if}
\State \textbf{else if} $t \geq T$ \textbf{do}:
        \State \quad Choose an action according to the learner agent’s policy
     \State \textbf{end if}
\State The DRL-based xApp acts based on the action recommended in the previous algorithm steps, \emph{$a_{t}$}
\State Allocate PRBs to the available slices according to \emph{$a_{t}$}
\State Execute scheduling within each slice
\State  Calculate reward $R$ using \eqref{eqn:rewardequation}
\State Update the learner agent’s policy based on the reward received every $\beta$ step
\State $t \leftarrow t + 1$
\State $\theta \leftarrow  \theta$ * $\nu$
\end{algorithmic}
\end{algorithm}

One or more source expert policies are first trained and fine-tuned as defined in Section \ref{flows}. Then, an expert policy is directly reused to guide the target policy of a learner DRL agent of a newly deployed xApp \cite{10172347, 9789336}. This should also happen when the xApp experiences a significant change in network conditions. The learner agent is configured to consult the expert policy and follow its recommended actions given a state. This happens for $T$ time steps, namely transfer duration. Meanwhile, the target learner policy is continuously updated based on the reward feedback the learner agent receives. The expert policy is deterministic and is not updated.

We employ the concept of transfer rate similar to \cite{8016642}. This gives the newly deployed agent the flexibility to not rely fully on the expert policy but also consult the learner policy being trained. This is particularly important since, although the expert policy is trained using real network data, the granularity of such data and its generality constraints make the expert policy limited. Having a transfer rate enables consulting both the source and target policies based on a parameter $\theta$ configured by the MNO where $\pi = (1 - \theta) \pi_{L} + \theta \pi_{E}$. If $\theta=1$, this indicates that the action recommended by the expert policy is always taken during the first $T$ time steps after deployment. Then, the actions recommended by the updated learner policy are followed afterward. However, a smaller or different decaying transfer rate can be configured to switch between the source expert and target learner policies during exploration. For instance, upon deploying the O-RAN's xApp, $\pi_{E}$ is expected to perform better than $\pi_{L}$. The newly deployed DRL agent's policy may be more uncertain than the expert policy given a specific network context. However, as time passes, $\pi_{L}$ gradually becomes more adapted to the real network environment compared to the source expert policy. Thus, a decaying transfer rate is proposed so that the target learner policy takes more control as it approaches $T$ time steps.

If more than one expert policy is used, policy reuse can be in the form of a weighted combination of these source policies. Here, for a given state, the xApp greedily picks the action with the highest reward, from all the available policies. This is referred to as the generalized policy improvement theorem for policy reuse of one or more source policies. It can be represented as follows \cite{10.5555/3294996.3295161}: 

\begin{theorem}[Generalized Policy Improvement] \label{eqn:reuseequation} Let $\left\{\pi_{i}\right\}_{i=1}^{n}$ be $n$ policies and let $\left\{\hat{Q}^{\pi_{i}}\right\}_{i=1}^{n}$ be their approximated action-value functions, s.t: $\left|Q^{\pi_{i}}(\kappa, a)-\hat{Q}^{\pi_{i}}(\kappa, a)\right| \leq \epsilon \forall \kappa \in$ $\mathcal{K}, a \in \mathcal{A}$, and $i \in[n]$. Define $\pi(\kappa)=\arg \max\limits_{a} \max\limits_{i} \hat{Q}^{\pi_{i}}(\kappa, a)$, then: $Q^{\pi}(\kappa, a) \geq \max\limits_{i} Q^{\pi_{i}}(\kappa, a)-\frac{2}{1-\lambda} \epsilon, \forall \kappa \in \mathcal{K}, a \in \mathcal{A}$, where $\lambda$ is a discounted factor, $\lambda$ $\in (0,1]$.
\end{theorem}

\vspace{0.4em}

\subsubsection{Policy Distillation}

In policy distillation, one or more source policies are used to guide a target learner policy. This is done by minimizing the divergence of action distributions between the source expert policy $\pi_{E}$ and target learner policy $\pi_{L}$, which can be written as $\mathcal{H}^{\times}\left(\pi_{E}\left(\tau_{t}\right) \mid \pi_{L}\left(\tau_{t}\right)\right)$ \cite{10172347}:

\vspace{-0.6em}

\begin{equation}
\min _{L} \mathbb{E}_{\tau \sim \pi_{E}}\left[\sum_{t=1}^{|\tau|} \nabla_{L} \mathcal{H}^{\times}\left(\pi_{E}\left(\tau_{t}\right) \mid \pi_{L}\left(\tau_{t}\right)\right)\right]
\end{equation}

where this reflects an expectation that is taken over trajectories, $\tau$, sampled from the source expert policy $\pi_{E}$. In expert distillation approaches, $N$ expert policies are individually learned for $N$ source tasks. Consequently, each expert policy results in a dataset $D^{E}=\left\{\kappa_{i}, \boldsymbol{q}_{i}\right\}_{i=0}^{N}$. Such datasets are mainly comprised of states $\kappa$ and action values $\boldsymbol{q}$, such that 
\begin{equation}
\boldsymbol{q}_{i}=\left[Q\left(\kappa_{i}, a_{1}\right), Q\left(\kappa_{i}, a_{2}\right), \ldots \mid a_{j} \in \mathcal{A}\right]   
\end{equation}

Finally, expert policies should be distilled into one policy. As mentioned before, this can be done by minimizing the divergence between each expert policy $\pi_{E_{i}}(a \mid \kappa)$ and the learner policy $\pi_{L}$. One example is the KL-divergence that can be calculated as follows given the dataset $D^{E}$ \cite{rusu2015policy}: 

\vspace{-1em}

\begin{equation}
\Scale[0.935]{
    \min _{L} \mathcal{D}_{K L}\left(\pi^{E} \mid \pi_{L}\right) \\ \approx \sum_{i=1}^{\left|D^{E}\right|}\operatorname{softmax}\left(\frac{\boldsymbol{q}_{i}^{E}}{\tau}\right) \ln \left(\frac{\operatorname{softmax}\left(\boldsymbol{q}_{i}^{E}\right)}{\operatorname{softmax}\left(\boldsymbol{q}_{i}^{L}\right)}\right)}
\end{equation}

We are using one expert policy at a time in the slicing xApp scenario. Hence, we follow a similar approach by calculating a vector value exactly at the midpoint between the actions recommended by the expert policy $\pi_{E}$ and the learner policy $\pi_{L}$ given a state. Then, an action with the shortest Euclidean distance to that vector value is chosen from the action space as described in Algorithm \ref{alg:two} as follows:

\begin{algorithm}
\caption{Proposed Policy Distillation Approach} 
\label{alg:two}
\textbf{Input:} $\pi_{E}$, MNO configurations, $\kappa$\\
\textbf{Parameters:} $\theta$, $T$, $\Omega$, buffer size, $\beta$, transfer rate decay, $\nu$\\
\textbf{Output:} PRB Allocation per slice
\begin{algorithmic}[1]
\State Load the appropriate pre-trained expert policy from $\mathcal{P}_{\text{E}}$
\State Initialize the learner action value function with random weights or from a policy pre-trained using a significantly different traffic pattern
\State \textbf{if} $t < T$ \textbf{do}:
        \State\quad  Generate a random number $x$, where $0 \leq x \leq 1$
        \State \quad \textbf{if}{ $x$ $\leq$ $\theta$} \textbf{do}:
            \State \quad \quad Consult the expert policy
            \State  \quad   \quad Choose an action according to Theorem \eqref{eqn:reuseequation}
            \State \quad \quad Consult the learner agent’s policy
            \State  \quad  \quad Find the midpoint between the actions recommended \NoNumber{\quad\quad by the expert and learner policies}
            \State  \quad  \quad Calculate the Euclidean distance between such a  \NoNumber{\quad\quad vector and all actions in the action space according}  \NoNumber{\quad\quad to \eqref{eqn:euclideandistance} to get the closest action}
        \State \quad \textbf{else if} $x$ $>$ $\theta$ \textbf{do}:
            \State \quad \quad Choose an action according to the learner policy
        \State \quad \textbf{end if}
    \State \textbf{else if} $t \geq T$ \textbf{do}:
        \State \quad Choose an action according to the learner agent’s policy
    \State \textbf{end if}

\State The DRL-based xApp acts based on the action recommended in the previous algorithm steps, \emph{$a_{t}$}
\State Allocate PRBs to the available slices according to \emph{$a_{t}$}
\State Execute scheduling within each slice
\State  Calculate reward $R$ using \eqref{eqn:rewardequation}
\State Update the learner agent’s policy based on the reward received every $\beta$ step
\State $t \leftarrow t + 1$
\State $\theta \leftarrow  \theta$ * $\nu$
\end{algorithmic}
\end{algorithm}

\begin{figure*}
	\centering
	\includegraphics[width=0.85\linewidth]{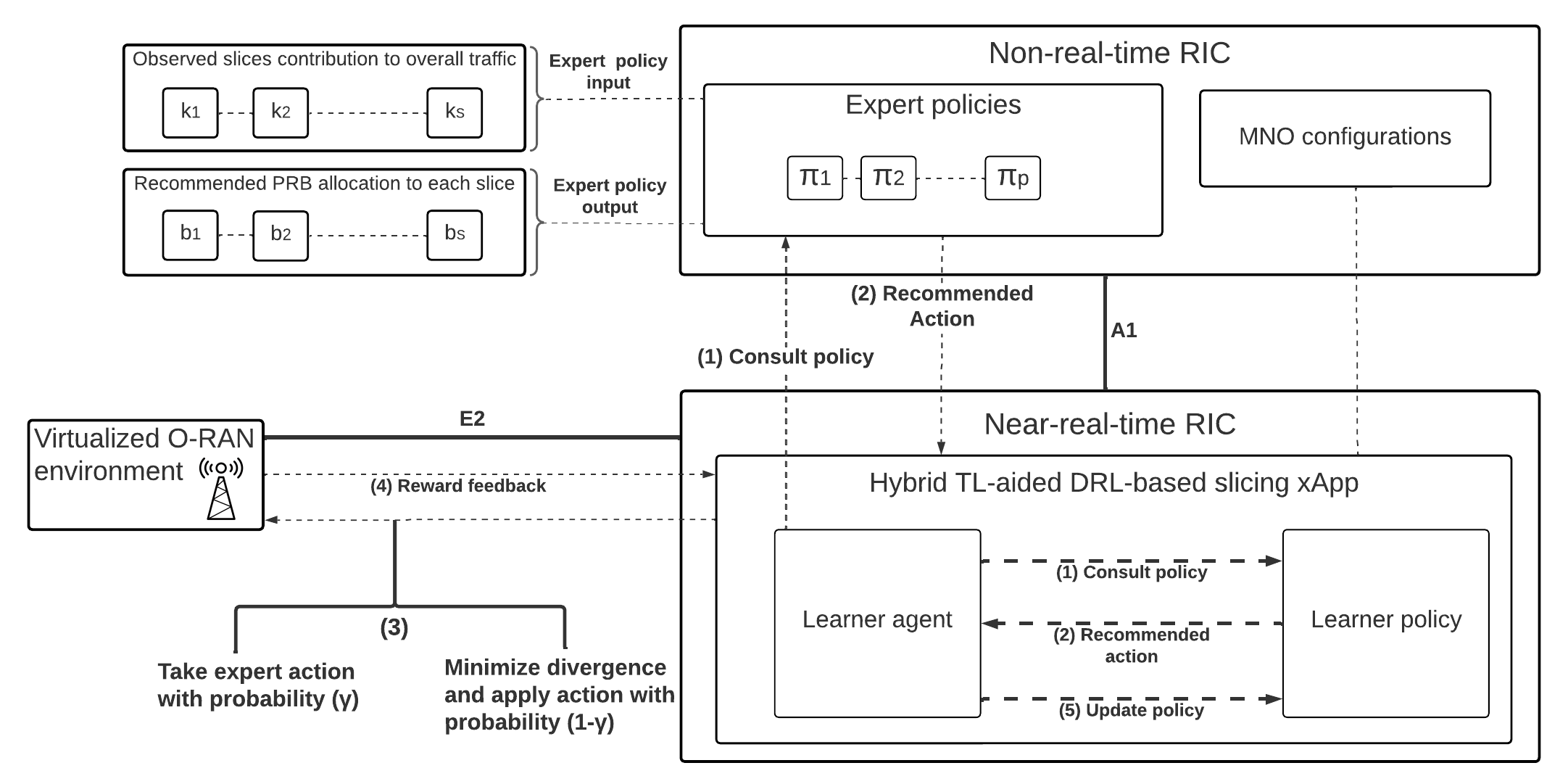}
  \setlength{\belowcaptionskip}{-12pt} 
	\caption{Block diagram of the proposed hybrid approach's components and interactions.}
	\label{hybrid-interactions}
\end{figure*}

\begin{equation}
\centering
\label{eqn:euclideandistance}
{ 
d \left( {a_{\pi_{L}}},{a_{\pi_{E}}}\right)   = \sqrt {\sum _{s=1}^{S}  \left( {a_{\pi_{L}}}_{s}-{a_{\pi_{E}}}_{s}\right)^2 }
}
\end{equation}

where $a_{\pi_{E}}$ and $a_{\pi_{L}}$ are vectors of actions recommended by the expert policy and the agent's learner policy respectively. Again, the learner agent of the deployed xApp follows the distilled policy with a probability that depends on the transfer rate, $\theta$, configured by the MNO.

\vspace{-1em}

\subsection{Proposed Hybrid Policy Transfer Approach}
\label{sec:hybrid}

Employing transfer learning should generally result in gains when compared with the non-TL-aided DRL approach. The knowledge of expert policies fine-tuned in a live network is reused to guide the learner agent instead of randomly exploring the action space. The policy reuse approach, however, is expected to perform poorly when the expert policy is trained on traffic patterns that are very different from those of the actual deployment environment. Hence, policy reuse delays the DRL agent's recovery when there is a big discrepancy between the source domain and the target domain.

On the other hand, policy distillation may prevent the learner agent from gaining the maximum possible rewards in some cases. For instance, this may happen when it is being guided by an expert policy that was trained on traffic patterns that are very similar to those of the actual deployment environment. Hence, it is expected that the two aforementioned policy transfer approaches will have some drawbacks in certain situations. We propose a hybrid of the two approaches to achieve a more robust TL-aided DRL exploration and increase the overall reward feedback. This is helpful when the transferred policies are not generic enough to robustly adapt to new traffic patterns. We introduce a parameter, $\gamma$, similar to $\theta$ to balance between exploiting the expert policy and exploring a distilled action.

The proposed approach is implemented using the proposed deployment workflow that adheres to the O-RAN architecture. Guidance is carried out by modifying the exploration process \cite{10.5555/2789272.2886795}. This allows the learner agent to use an expert policy with probability $\gamma$, and minimize the divergence between the expert and learner policies with probability (1 - $\gamma$). The reused policy may have been learned under similar or different traffic conditions relative to the current conditions. Our proposed hybrid transfer learning approach combines two TL approaches to accommodate these two situations.
The first is policy reuse which directly follows the expert policy’s recommended action. This is beneficial when the reused policy is pre-trained under similar conditions to that of the learner agent. The second is policy distillation, which minimizes the divergence between the expert and the learner policies’ actions. This is beneficial when the reused policy is pre-trained under different conditions from that of the learner agent. A hybrid approach allows the DRL agent to start with a good reward value whenever it is newly deployed in a live network. It also enables the agent to converge quickly to the optimal slicing configuration compared with the two approaches separately. 

The main components of the proposed hybrid approach and the interactions between its components are visualized in Fig. \ref{hybrid-interactions}. As summarized in Algorithm \ref{alg:three}, the proposed approach follows the steps below:

\begin{algorithm}
\caption{Proposed Hybrid Policy Transfer Approach} 
\label{alg:three}
\textbf{Input:} $\pi_{E}$, MNO configurations, $\kappa$\\
\textbf{Parameters:} $\theta$, $T$, $\Omega$, $\gamma$, buffer size, $\beta$, transfer rate decay, $\nu$\\
\textbf{Output:} PRB Allocation per slice
\begin{algorithmic}[1]
\State Load the appropriate pre-trained expert policy from the stored policies $\mathcal{P}_{\text{E}}$
\State Initialize the learner action value function with random weights or from a policy pre-trained using a significantly different traffic pattern
\State \textbf{if} $t < T$ \textbf{do}:
        \State\quad  Generate a random number $x$, where $0 \leq x \leq 1$
        \State \quad \textbf{if}{ $x$ $\leq$ $\theta$} \textbf{do}:
        \State \quad \quad Generate a random number $r$, where $0 \leq r \leq 1$
            \State \quad \quad  \textbf{if}{ $r$ $<$ $\gamma$} \textbf{do}:
            \State \quad \quad  \quad Consult the expert policy
            \State  \quad \quad \quad Choose an action according to Theorem \eqref{eqn:reuseequation}
        \State \quad \quad \textbf{else if}{ $r$ $\geq$ $\gamma$} \textbf{do}:
            \State \quad \quad  \quad Consult the learner policy
            \State  \quad   \quad \quad Find the midpoint between the actions recom-\NoNumber{\quad\quad\quad mended by the expert and learner policies}
            \State  \quad   \quad  \quad Calculate the Euclidean distance between such a\NoNumber{\quad\quad\quad vector and all actions in the action space according}\NoNumber{\quad\quad\quad to \eqref{eqn:euclideandistance} to get the closest action}
     \State \quad \quad  \textbf{end if}
        \State \quad \textbf{else if} $x$ $>$ $\theta$ \textbf{do}:
            \State \quad \quad Choose an action according to the learner policy
        \State \quad \textbf{end if}
    \State \textbf{else if} $t \geq T$ \textbf{do}:
        \State \quad Choose an action according to the learner agent’s policy
    \State \textbf{end if}

\State The DRL-based xApp acts based on the action recommended in the previous algorithm steps, \emph{$a_{t}$}
\State Allocate PRBs to the available slices according to \emph{$a_{t}$}
\State Execute scheduling within each slice
\State  Calculate reward $R$ using \eqref{eqn:rewardequation}
\State Update the learner agent’s policy based on the reward received every $\beta$ step
\State $t \leftarrow t + 1$
\State $\theta \leftarrow  \theta$ * $\nu$
\end{algorithmic}
\end{algorithm}

\begin{table*}
\centering
\caption{Experiment Setup: RAN slicing DRL design.}
\begin{tabular}{|p{1.in}|p{1in}|p{1in}|p{1in}|}
\hline
\multicolumn{1}{|l|}{\textbf{State}}                                                                                                                        & \multicolumn{3}{l|}{\begin{tabular}[c]{@{}l@{}}Slices' contribution to the overall BS's traffic within preceding $\Omega$ as in (\ref{eq:state-represenation})\end{tabular}}                                                     \\ \hline
\multicolumn{1}{|l|}{\textbf{Action}}                                                                                                                       & \multicolumn{3}{l|}{\begin{tabular}[c]{@{}l@{}}PRBs allocated to each slice as defined in (\ref{eq:action})\end{tabular}} \\ \hline
\multicolumn{1}{|l|}{\textbf{Reward function}}                                                                                                                       & \multicolumn{3}{l|}{\begin{tabular}[c]{@{}l@{}}A weighted sum of a sigmoid function of the average latency experienced in a slicing \\window by the various slices as defined in (\ref{eqn:rewardequation})\end{tabular}} \\ \hline
\multicolumn{1}{|l|}{\textbf{Reward function weights}}                                                                                                                       & \multicolumn{3}{l|}{$w_{\text{VoNR}}=0.1$, $w_{\text{VR gaming}}=0.7$, $w_{\text{Video}}=0.2$}                                                                                                                                                                \\ \hline
\multicolumn{1}{|l|}{\textbf{Reward function parameters}}                                                                                                                       & \multicolumn{3}{l|}{${c_{2}}_{\text{VoNR}} = 10 $, ${c_{2}}_{\text{VR gaming}} =  1$, ${c_{2}}_{\text{Video}} =  5$}                                                                                                                                                                \\ \hline
\multicolumn{1}{|l|}{\textbf{DRL algorithm}}                                                                                                                       & \multicolumn{3}{l|}{Proximal Policy Optimization (PPO)}                                                                                                                                                                \\ \hline
\multicolumn{1}{|l|}{\textbf{Learning steps per run}}                                                                                                                       & \multicolumn{3}{l|}{10,000}                                                                                                                                                                \\ \hline
\multicolumn{1}{|l|}{\textbf{Exploration rates}}                                                                                                                       & \multicolumn{3}{l|}{0.2, exploration ends at step 4000}                                                                                                                                                                \\ \hline
\multicolumn{1}{|l|}{\textbf{Exploration decay rates}}                                                                                                                       & \multicolumn{3}{l|}{0.99, 0.7, 0.5, 0.3       }                                                                                                                                                         \\ \hline
\multicolumn{1}{|l|}{\textbf{Transfer rate}}                                                                                                                       & \multicolumn{3}{l|}{0.9, 0.7, 0.5, 0.3, $T = 3000$}                                                                                                                                                                \\ \hline
\multicolumn{1}{|l|}{\textbf{Hybrid approach parameter}}                                                                                                                       & \multicolumn{3}{l|}{0.99, 0.9, 0.7, 0.5, 0.3}                                                                                                                                                               \\ \hline
\multicolumn{1}{|l|}{\textbf{Learning rate}}                                                                                                                       & \multicolumn{3}{l|}{0.01}                                                                                                                                                                \\ \hline
\multicolumn{1}{|l|}{\textbf{Batch size}}                                                                                                                       & \multicolumn{3}{l|}{4}                                                                                                                                                                \\ \hline

\end{tabular}
\label{tab:drl_parameters}
\end{table*}

\begin{enumerate}

\item The slicing xApp consults both the expert and the learner policies to get their recommended actions given the current system state.
\item The expert and the learner policies provide the xApp with their recommended actions to take. 
\item The learner agent of the slicing xApp decides whether to follow its own policy, the expert policy, or a distilled policy. This is mainly determined by the configuration of $\gamma$ and $\theta$ set by the MNO as described in Algorithm \ref{alg:three}. A bigger value of $\gamma$ implies that the distillation will happen less often during the transfer time $T$. Based on that decision, the slicing xApp executes the proper action to allocate PRBs among the admitted slices. 
\item The slicing xApp logs the relevant KPIs and calculates the reward feedback based on the reward function defined by the MNO. 
\item The slicing xApp updates the learner policy based on the received reward. The frequency of such updates depends on parameters such as the buffer size. Moreover, the policy update equation depends on the DRL algorithm followed by the learner agent. The expert policy is deterministic and cannot be updated.
\item After $T$ time steps, the xApp follows the latest version of the learner policy until experiencing significant changes in the network conditions. This can be detected in many ways but it is not the focus of this paper. For instance, the TL-aided DRL slicing xApp can track the reward feedback it receives. It can then start a new policy transfer procedure when the reward values become lower than a threshold defined by the MNO.

\end{enumerate}

\begin{table*}
\centering
\caption{Experiment Setup: Simulation Parameters Settings}
\begin{tabular}{|p{1.15in}|p{1.6in}|p{1.6in}|p{1.6in}|}
\hline
\textbf{\begin{tabular}[c]{@{}l@{}} \end{tabular}}   & \textbf{Video}     & \textbf{VoNR}                & \textbf{VR gaming}                                                                                                  \\ \hline

\multicolumn{1}{|l|}{\textbf{Scheduling algorithm}}                                                                                                                        & \multicolumn{3}{l|}{\begin{tabular}[c]{@{}l@{}}Round-robin per 1 ms slot\end{tabular}}                                                     \\ \hline
\multicolumn{1}{|l|}{\textbf{Slicing window size}}                                                                                                                       & \multicolumn{3}{l|}{\begin{tabular}[c]{@{}l@{}}PRB allocation among slices every 100 scheduling time slots\end{tabular}} \\ 

\hline
\textbf{\begin{tabular}[c]{@{}l@{}}Packet interarrival time \end{tabular}}   & Truncated Pareto (mean = 6 ms, max = 12.5 ms)      & Uniform (min = 0 ms, max = 160 ms)                                                       & Real VR gaming dataset \cite{9685808}                                                                                                     \\ \hline

\textbf{\begin{tabular}[c]{@{}l@{}}Packet size\end{tabular}}                & Truncated Pareto (mean = 100 B, max = 250 B) & Constant (40 B)                                                                    & Real VR gaming dataset \cite{9685808}                                                   \\ \hline
\textbf{\begin{tabular}[c]{@{}l@{}}Number of users\\ \end{tabular}}                & Poisson (max = 43, mean = 20)                       & Poisson (max = 104, mean = 70)                                                         & Poisson (max = 7, mean = 1)                                                          \\ \hline
\end{tabular}
\label{tab:sim_parameters}
\end{table*}

\section{Simulations and Results}
\label{sec:simulations}

\subsection{Simulation Settings}
\label{sim-setup}

We follow the DRL design described in Section \ref{mapping} and summarized in Table \ref{tab:drl_parameters}. We conduct a thorough study to test the proposed training and deployment O-RAN flows. We also examine the convergence performance of the three proposed policy transfer-aided DRL approaches in the context of O-RAN slicing. To do so, we follow an approach similar to the one proposed in Section \ref{flows}. We first train several expert models using real VR gaming traffic \cite{9627832} to reflect practical scenarios of immersive applications in 6G networks. The VR gaming data includes multiple games and multiple configurations per game. Additionally, voice over new radio (VoNR) and video requests are generated based on the parameters described in Table \ref{tab:sim_parameters} following models similar to the ones defined in \cite{8540003}. VR gaming users generate the largest requests. Moreover, video users receive packets more frequently compared to the other two service types. Finally, VoNR users generate small and constant-size requests.

We then save these various models in a policy directory as shown in Fig. \ref{TL-guided}. After that, we start the deployment process in which the saved policies are used to guide the DRL-based slicing xApp following the proposed flow. We assume that the DRL agent is newly deployed in an environment or that the context has just changed significantly. We also assume that an expert policy is already chosen and loaded at the beginning of our simulation runs to guide the learner agent. The learner policy is initialized randomly. Such a configuration is set to reflect the big difference between the offline training simulation and the real deployment environments. This also reflects a significant change in the network’s conditions, and hence, the need for exploration in both cases. This enables evaluation of the proposed approach’s performance against the baselines under extreme conditions. However, our framework can be configured to accommodate any pre-trained policy to be used as an initial policy for the learner DRL agent.

We compare the three proposed TL-aided DRL approaches with their traditional non-TL-aided DRL counterparts. This allows us to study the gains in convergence performance in terms of the average initial reward value, convergence rate, rewards variance per run, and the number of converged scenarios. This evaluates some of the safety and acceleration aspects of the proposed approaches. For that, we also use various traffic patterns from the VR gaming dataset to reflect two main learner agent scenarios:

\begin{enumerate}
    \item A scenario that includes DRL-based slicing xApps that are deployed in an environment experiencing traffic patterns similar to those used to train the expert policies guiding such xApps.
    \item A scenario in which the slicing xApps are experiencing traffic patterns that are different from the patterns used to train the expert policies that guide the policy transfer process.
\end{enumerate}

In both the training and deployment flows, the DRL-based slicing xApp allocates the limited PRBs to the available 3 slices. After that, round-robin scheduling is executed within each slice at the granularity of 1 ms. The slicing window size is 100 ms. Hence, scheduling continues for 100 time slots. The scheduler can allocate resources to multiple transmissions per transmission time interval (TTI) if enough resources are available. Moreover, unsatisfied users leave the system if they have several unfulfilled requests. The reward function defines the goal of an RL problem \cite{sutton2018reinforcement}. In our experiment, the goal is to keep the latency of the different slices in an acceptable range defined by the slices’ SLAs. This is reflected by the $c_{2}$ parameter in (\ref{eqn:rewardequation}). Thus, unsatisfied users are considered implicitly in terms of high latency before they leave. This enables the reward function to penalize the DRL agent for taking actions that lead to high delays during a given slicing window, and hence, users leaving the system.

We used the configurations defined in Table \ref{tab:drl_parameters} for all the expert and learner agents. For better generality of the results, we ran all the possible combinations out of the listed hyper-parameters of the implemented approaches. The agents considered in this paper employ the proximal policy optimization (PPO) algorithm \cite{schulman2017proximal} as an underlying DRL algorithm. We adopt the PPO implementation from the Tensorforce Python package \footnote{{Available at \url{https://github.com/tensorforce/tensorforce}}}. We modified it accordingly, to accommodate the flows and the policy transfer algorithms proposed in Section \ref{flows-section}. VR gaming slices are relatively more latency intolerant. Therefore, the reward function weights and parameters are configured to reflect such sensitivity as seen in Table \ref{tab:drl_parameters}. 

The O-RAN specifications mandate that any ML-based solution should not be trained online. It can only be fine-tuned online to ensure that the trained models do not affect the performance and stability of the network \cite{10071941}. To avoid that, we configure the DRL agent to have a low initial exploration rate. We also employ an exploration decay to restrict the random actions exploration of the learner agents. We study three primary aspects of the results. We start with analyzing the reward convergence behavior of the different approaches. We then evaluate their safety and acceleration aspects. We finally investigate the effect of the introduced parameter $\gamma$ on the performance of the proposed hybrid TL-aided DRL approach.

All the experiments carried out in this study were conducted on a Linux machine with 8 CPUs, 64 GB of RAM, and an NVIDIA GeForce RTX 2080Ti GPU. The developed simulation environment that includes the implementation of the proposed methods is available on GitHub\footnote{{Available at \url{https://github.com/ahmadnagib/TL-aided-DRL}}}. Such a DRL environment follows OpenAI Gym standards\footnote{\href{https://www.gymlibrary.dev/}{OpenAI Gym: https://www.gymlibrary.dev/}}. This allows the development of methods that can interact instantly with the environment. Hence, enables fellow researchers to reuse, extend, and compare the proposed approaches against their approaches.

\subsection{Reward Convergence Behaviour}

\begin{figure*}[ht] 
    
  \begin{subfigure}[b]{0.5\linewidth}
    \centering
\includegraphics[width=0.95\linewidth]{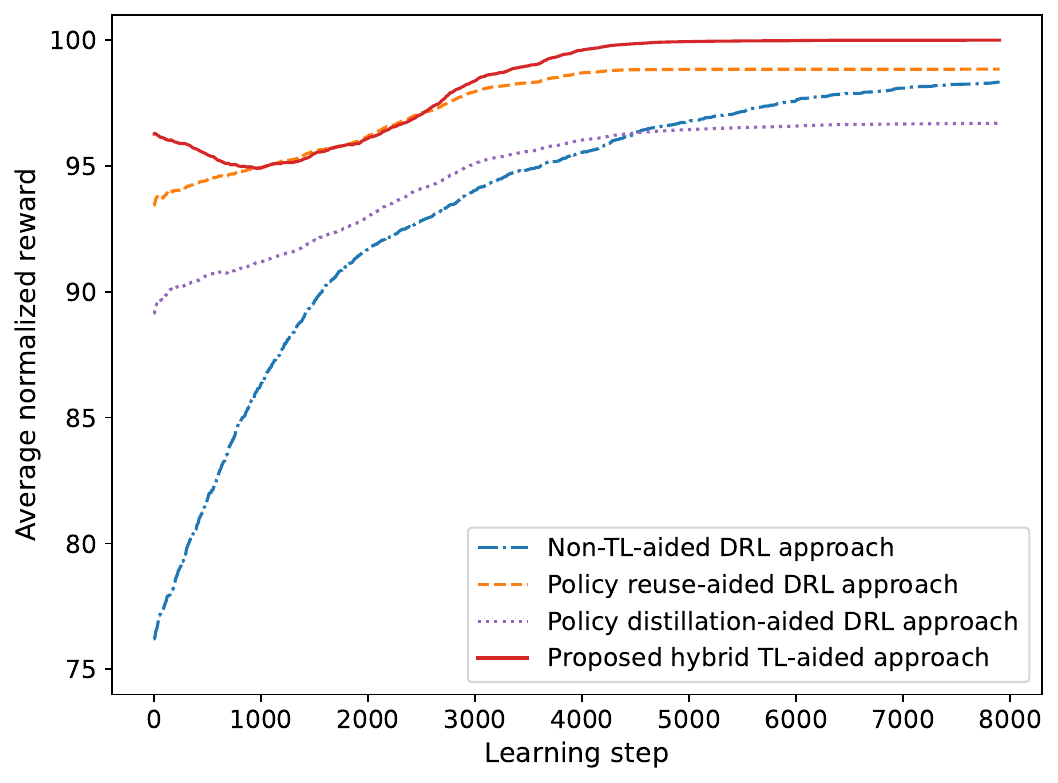} 
    \caption{} 
    \label{fig5:a} 
  \end{subfigure}
  \begin{subfigure}[b]{0.5\linewidth}
    \centering
    \includegraphics[width=0.95\linewidth]{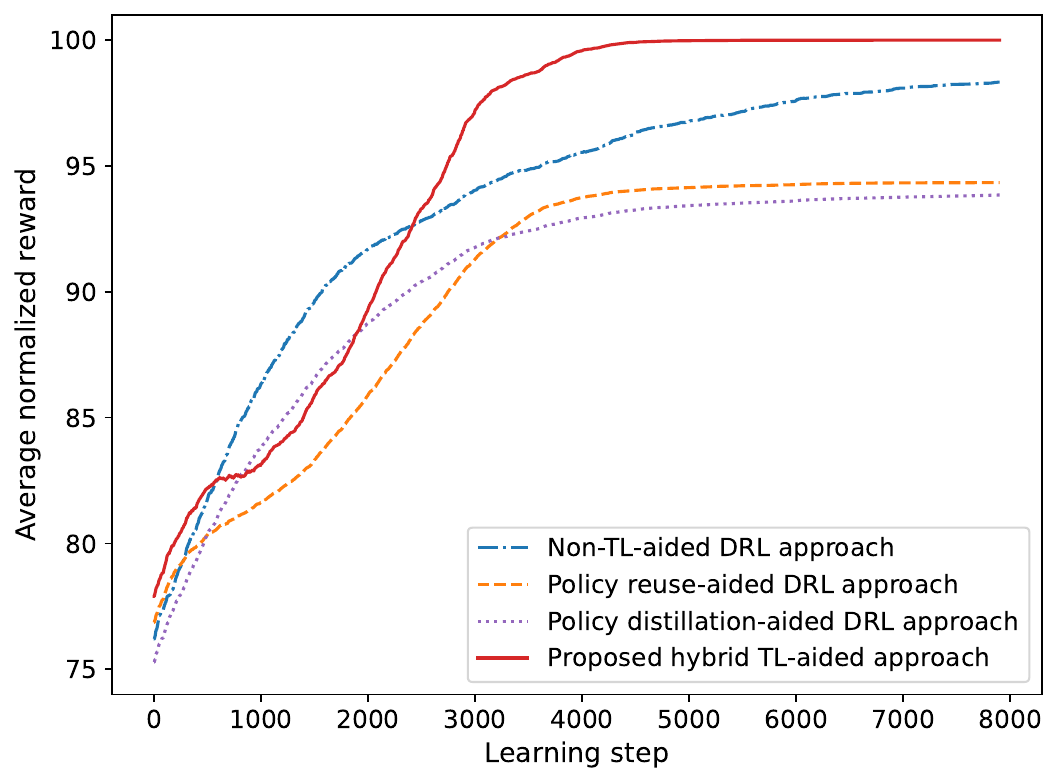} 
    \caption{} 
    \label{fig5:b}
  \end{subfigure} 
  \begin{subfigure}[b]{0.5\linewidth}
    \centering
    \includegraphics[width=0.95\linewidth]{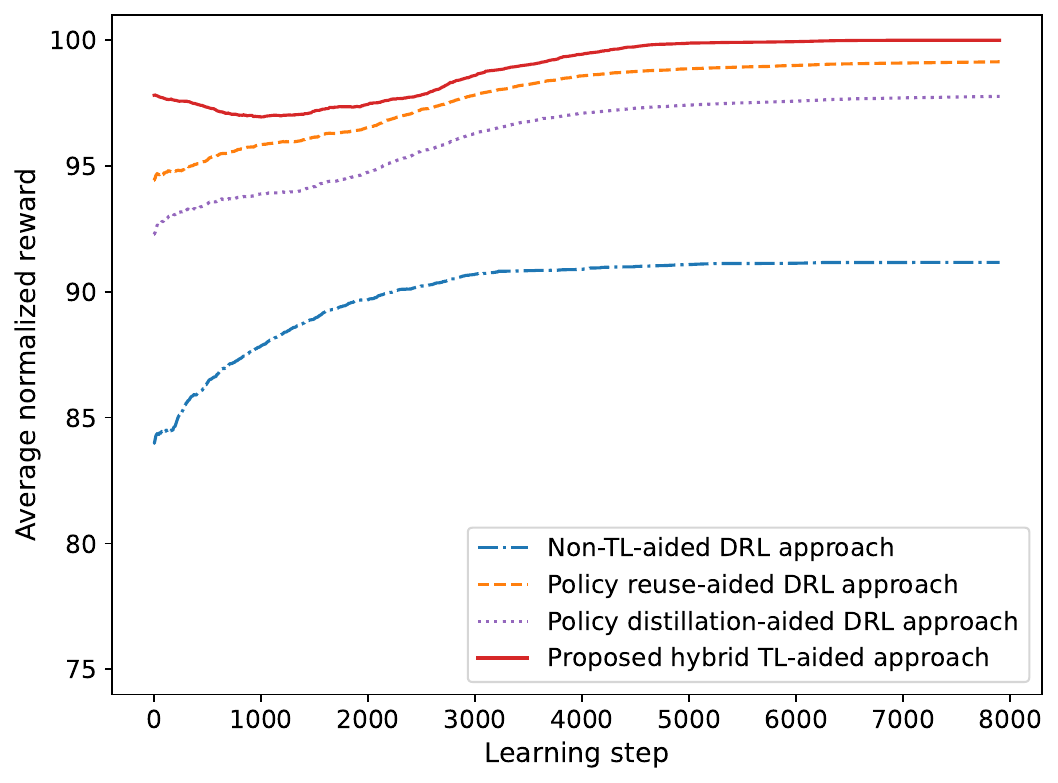} 
    \caption{} 
    \label{fig5:c} 
  \end{subfigure}
  \begin{subfigure}[b]{0.5\linewidth}
    \centering
    \includegraphics[width=0.95\linewidth]{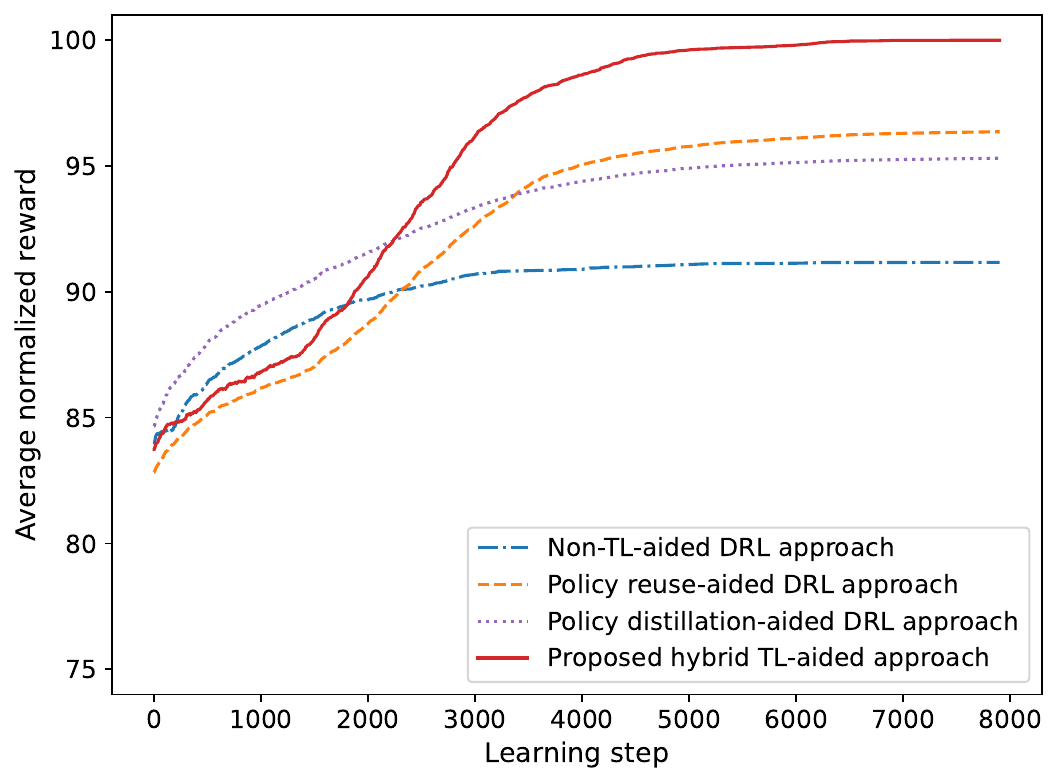} 
    \caption{} 
    \label{fig5:d} 
  \end{subfigure} 
     \setlength{\belowcaptionskip}{-12pt} 
  \caption{Reward convergence of the proposed approaches: a) and c) traffic patterns 1 and 2 guided by an expert policy trained using a similar traffic pattern; b) and d) traffic patterns 1 and 2 guided by an expert policy trained using a different traffic pattern (average of best 64 runs).}
  \label{fig:results1}

\end{figure*}

It is better to reuse local expert policies that were trained in contexts similar to the deployment environment. However, these are not always available. Hence, we show the reward convergence performance of the proposed approaches when an expert policy trained in a similar or different context is reused. As described in Section \ref{sim-setup}, we run all the combinations in Table \ref{tab:drl_parameters}. We choose the best 64 runs of each approach in terms of the average normalized reward per run and show the average rewards over the duration of a simulation run in Fig. \ref{fig:results1}.

Fig. \ref{fig5:a} and Fig. \ref{fig5:c} show the average convergence performance of the approaches in scenarios where the expert policy is trained using a traffic pattern similar to the deployment environment pattern. The hybrid approach has the highest initial reward value in both cases. The policy reuse approach comes second and its convergence behavior is very similar to that of the hybrid approach. This is primarily attributed to the similarity between the source and target policies' environments. The policy reuse follows the expert policy's actions and they are of high return most of the time due to such similarity. Nevertheless, unlike the other approaches, the hybrid approach accommodates the small differences between training and deployment environments by additionally following a distilled action for some time during the transfer time, depending on the $\gamma$ parameter setting. This enables it to explore in a safer way. Hence it converges to the best average reward in both situations.

\begin{figure*}[ht] 
  \begin{subfigure}[b]{1\linewidth}
    \centering    \includegraphics[width=0.65\linewidth]{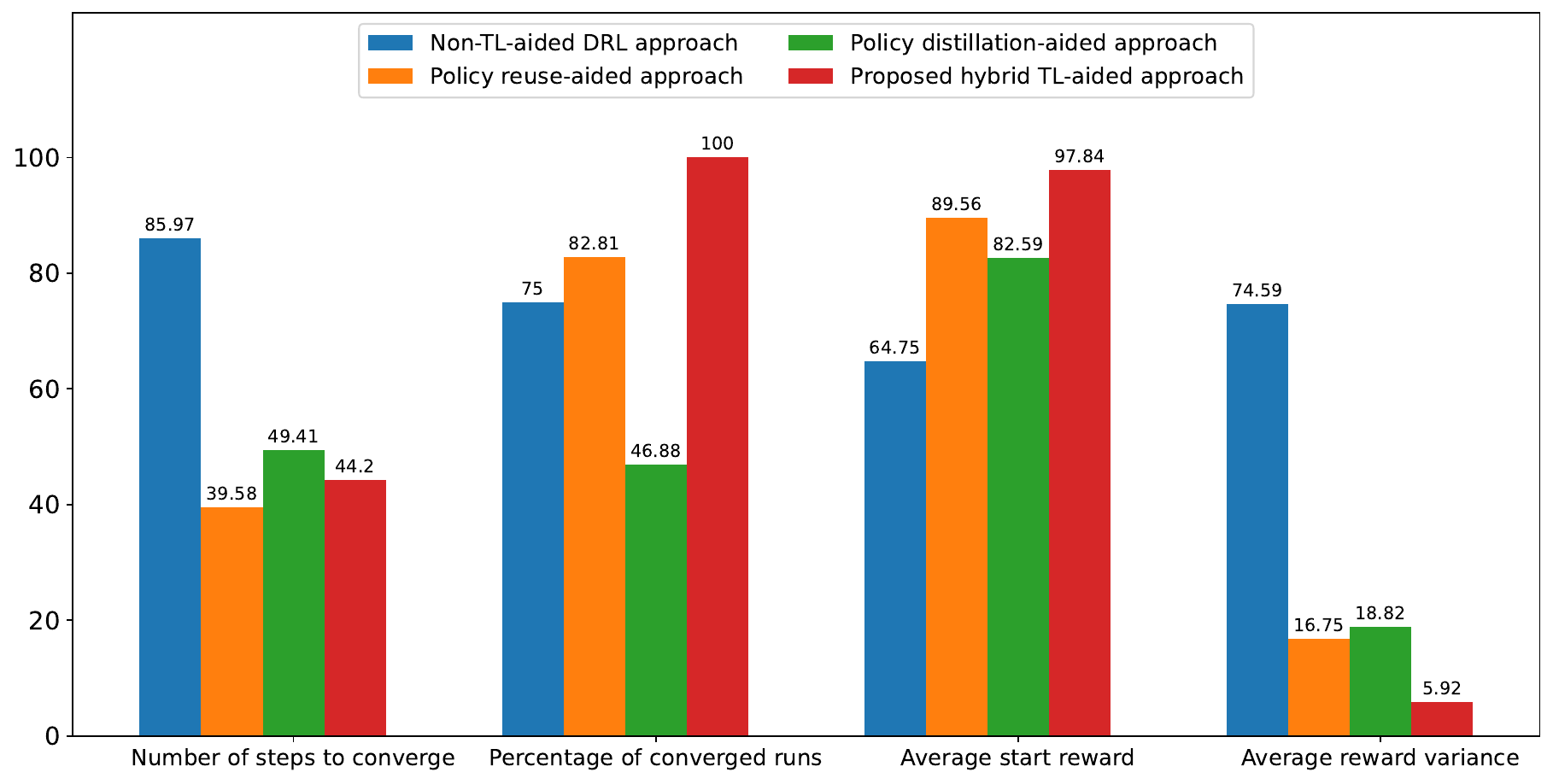} 
    \caption{} 
    \label{fig7:a} 
  \end{subfigure}
  
  \begin{subfigure}[b]{1\linewidth}
    \centering    \includegraphics[width=0.65\linewidth]{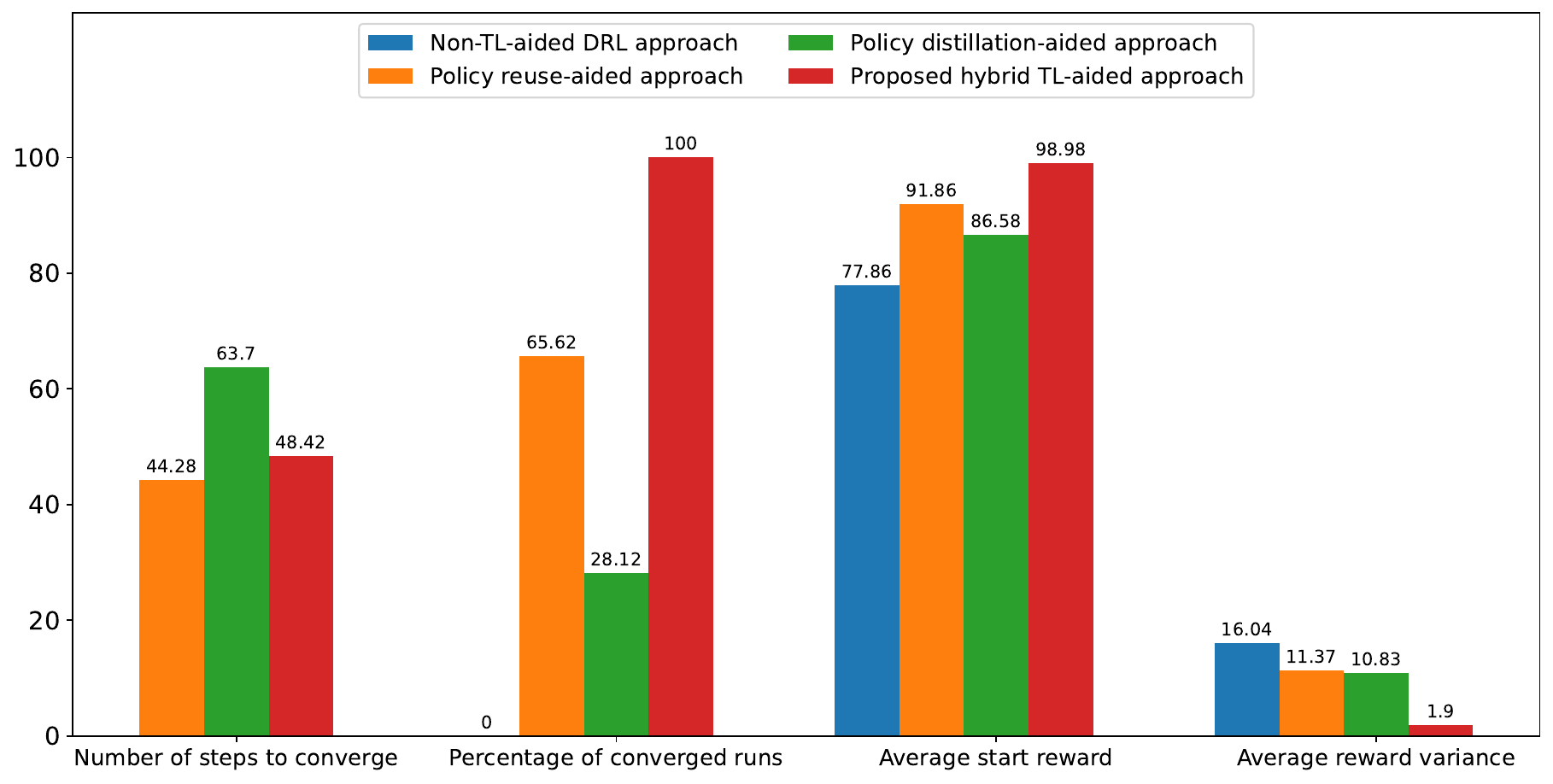} 
    \caption{} 
    \label{fig7:b} 
  \end{subfigure} 
 \setlength{\belowcaptionskip}{-12pt}   
  \caption{Safety and acceleration performance of the proposed approaches averaged over 64 best runs (the higher the better for the percentage of converged runs and the average start reward): a) traffic pattern 1; b) traffic pattern 2.}
  \label{fig:results3}
\end{figure*}

On the other hand, Fig. \ref{fig5:b} and Fig. \ref{fig5:d} show the average performance of the approaches in scenarios where the expert policy is trained using a different traffic pattern. The proposed approach still has the best overall reward convergence performance. The policy reuse approach, however, has almost the worst start and average reward values for a significant percentage of the simulation run duration. This can be attributed to the differences between the source and target policies environment. Hence, blindly following the expert policy given the restricted exploration will not lead to optimal actions as the network conditions are different. The policy distillation approach has a much better start, however, it converges to a sub-optimal value function as it tries to explore very carefully by finding an action that minimizes the divergence between the expert and the learner policies' actions. This prevents policy distillation from exploring the other possible high-return actions before the end of the limited 4000 exploration steps. Again, the hybrid approach accommodates the differences between training and deployment environments by switching between an expert policy reuse action and a distilled action depending on the $\gamma$ parameter setting.

The non-hybrid approaches are not able to explore the whole action space given the restricted exploration setting in terms of initial exploration, exploration decay, and exploration end step defined in Table \ref{tab:drl_parameters}. Consequently, they sometimes fail to converge to the optimal slicing configurations.  On the other hand, the DRL agents following the hybrid approach are guided by two kinds of live network knowledge. The exploration process of the hybrid approach is modified to incorporate such knowledge. Thus, it does not require as much random exploration or following the local learner policy as the other approaches during the transfer time, $T$.

\subsection{Safety and Acceleration Evaluation}

We now present statistics compiled from the best 64 runs of all the approaches given traffic patterns 1 and 2 in Fig. \ref{fig7:a} and Fig. \ref{fig7:b} respectively. The figure depicts the acceleration and safety aspects of the different approaches. More specifically, we measure the initial average normalized reward, variance in the reward, number of steps to converge to the best reward, and percentage of converged simulation runs for each approach. This measures whether an approach starts with a good reward value, the change in reward values afterward, the speed of convergence, and the ability to finally converge to the optimal policy respectively.

Such observations confirm the results presented in Fig. \ref{fig:results1} and the hypothesis made in Section \ref{sec:hybrid}. The proposed hybrid approach tries to maximize the reward for some time. It also tries to cautiously explore new actions by striking a balance between a deterministic and a guided exploratory action at other times. It inherits the best of both policy reuse and distillation approaches regardless of the nature of the expert policy's training environment and the actual deployment traffic conditions.

\begin{figure*}[ht] 
  \begin{subfigure}[b]{1\linewidth}
    \centering    \includegraphics[width=0.68\linewidth]{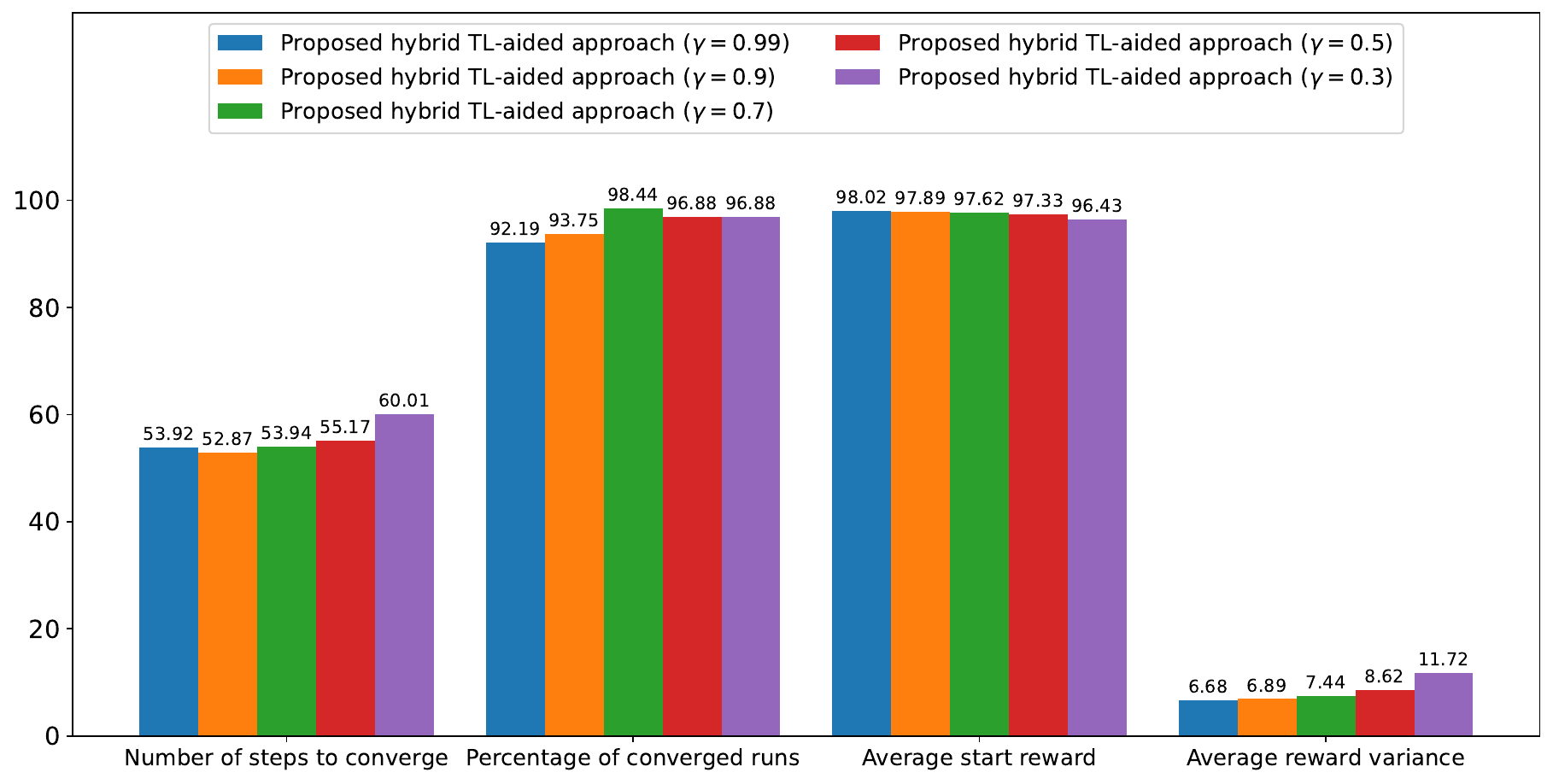} 
    \caption{} 
    \label{fig8:a} 
  \end{subfigure}

  \begin{subfigure}[b]{1\linewidth}
    \centering
    \includegraphics[width=0.68\linewidth]{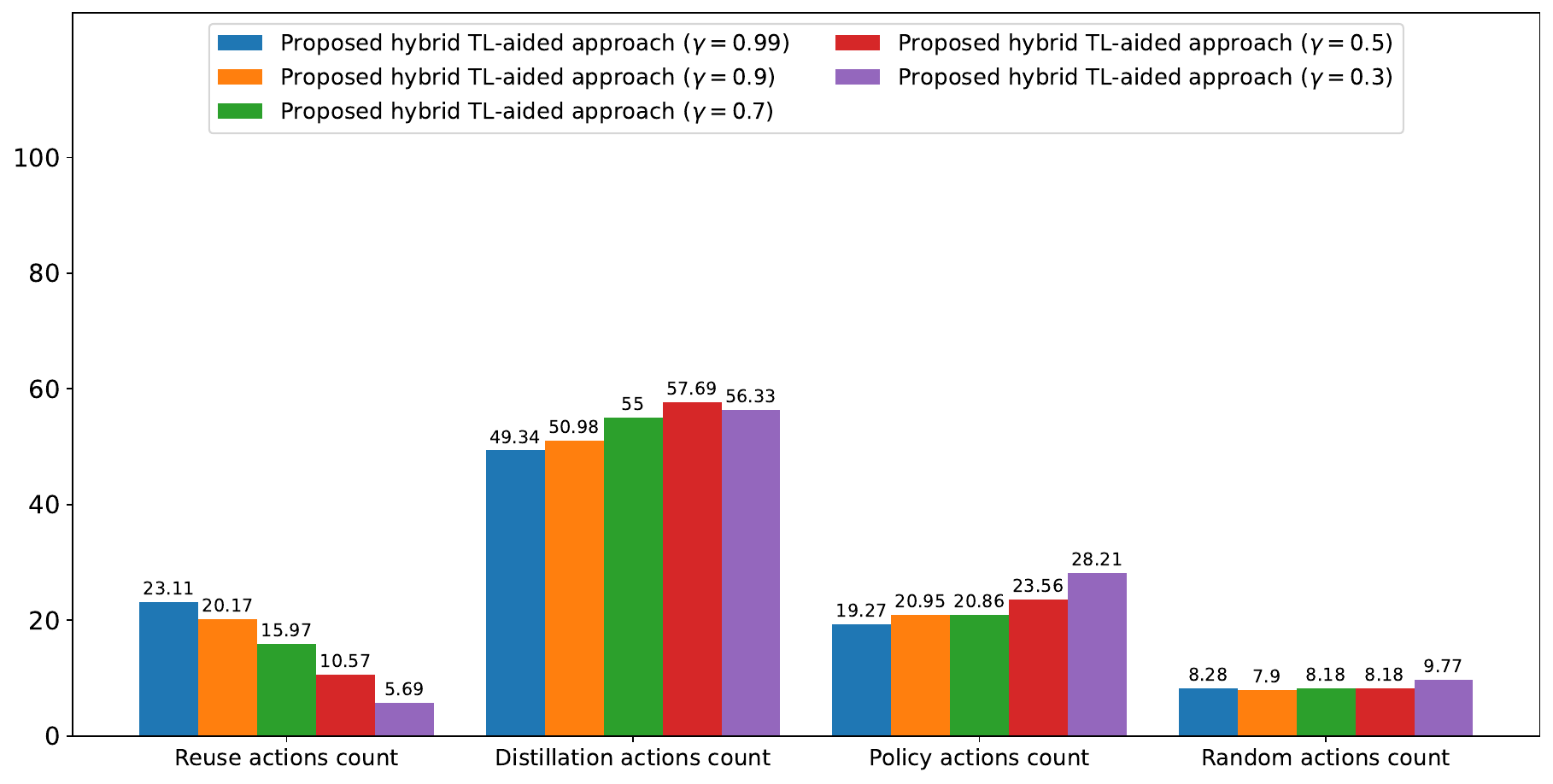}
    \caption{} 
    \label{fig8:b} 
  \end{subfigure} 
   \setlength{\belowcaptionskip}{-12pt} 
  \caption{The effect of the introduced parameter $\gamma$ on the convergence performance averaged over 64 best runs given traffic pattern 2: a) safety and acceleration performance given different $\gamma$ values; b) action counts during transfer time $T$.}
  \label{fig:results4}
\end{figure*}

Consequently, the proposed hybrid approach has the highest initial reward value and the highest percentage of converged runs with at least 7.7\% and 20.7\% improvements over the policy reuse approach respectively. It also yields the lowest variance in reward values per run with at least a 64.6\% decrease in variance when compared with policy reuse. It does so while still having the second-best performance in terms of the number of steps to converge. However, policy reuse which comes first in this metric only converges up to 82.8\% of the time. Hence, the number of steps to converge is averaged over a smaller number of data samples.

The hybrid approach can switch between two approaches of knowledge transfer. This enables it to deal with various expert policies’ pre-training conditions whether they are similar or different from the experienced live network conditions. The non-TL-aided approach has no sources of knowledge. It only relies on its own policy and random exploration which is restricted in such an O-RAN deployment scenario. Thus, it has almost the worst performance on all the compared aspects except for the percentage of converged runs in the scenario presented in Fig. \ref{fig7:a}. The policy reuse and distillation are close in their overall performance except for the percentage of convergence runs. This is primarily because the two traffic patterns are not hugely different from those used to pre-train the expert policy. Thus, the policy reuse can manage to converge more frequently if it only relies on following the expert policy, unlike policy distillation which fails more often as it conservatively converges towards a much lower average reward.

\subsection{Effect of the Introduced Hybrid Transfer Learning Parameter}

We also examine the effect of the introduced parameter $\gamma$ as shown in Fig. \ref{fig:results4}. The two sub-figures show the average performance of the proposed approach based on the best 64 runs for each value of $\gamma$. Fig \ref{fig8:a} shows that the number of steps needed by a learner agent to converge increases with the decrease of $\gamma$ value. The hybrid TL-aided approach with $\gamma = 0.3$ needs around 60\% of the simulation run steps to converge to the optimal reward value. It also shows a slight increase in the reward variance per run and a slight decrease in the initial reward value. This can be attributed to fewer policy reuse-based actions taken as defined in Algorithm \ref{alg:three} and showcased in Fig. \ref{fig8:b}. The used traffic pattern is still not very different from that used to pre-train the expert policies. The policy reuse approach showed a slight advantage over policy distillation in such situations as in Fig. \ref{fig:results3} and Fig. \ref{fig:results1}. Hence, an overall slight degradation in performance is expected. However, the hybrid approach still shows robust behavior given the different $\gamma$ values. This is due to restricting random actions and relying on both distillation and reuse during the majority of the transfer time as seen in Fig. \ref{fig8:b}. It is worth noting that the probability of taking a reuse or distillation action does not only rely on $\gamma$ and hence their counts do not change identically when changing $\gamma$.

\section{Conclusion and Future Work}
\label{conclusions}

Reusing existing knowledge is a major step towards having \textit{safe and accelerated} DRL-based xApps in the O-RAN paradigm. In this paper, we propose a hybrid TL-aided DRL approach that combines policy reuse and distillation TL methods. A thorough study on intelligent O-RAN slicing is conducted to demonstrate the DRL convergence performance gains of using the proposed approach.
For this, a public VR cloud gaming dataset is incorporated to reflect an example of realistic immersive applications of O-RAN slicing. The proposed hybrid approach proves to be effective whether the expert policies are pre-trained in a context similar to or different from that of the deployment environment. Results show at least: 7.7\% and 20.7\% improvements in the average initial reward value and the number of converged scenarios, and a 64.6\% decrease in reward variance while maintaining fast convergence and enhancing the generalizability compared with the baselines. This facilitates a \textit{safe and accelerated} DRL convergence when a slicing xApp is newly deployed in a live network and when the network context changes significantly.

Although the proposed hybrid approach proves to outperform the baselines, the associated hyper-parameters need dynamic optimization based on the context of both the deployment and expert policy training environments. Studying how to conditionally trigger policy transfer instead of relying on probabilities during the transfer time is another interesting research problem that should be addressed. Furthermore, research about the benefits and ways to reuse imperfect and low-cost policies is needed. Finally, combining the proposed method with other approaches such as constrained DRL \cite{10.5555/2789272.2886795} and time series forecasting \cite{nagib2023does} is a promising step toward trustworthy DRL in O-RAN slicing.

\bibliographystyle{IEEEtran}
\bibliography{references.bib}

\begin{thebibliography}{10}
\providecommand{\url}[1]{#1}
\csname url@samestyle\endcsname
\providecommand{\newblock}{\relax}
\providecommand{\bibinfo}[2]{#2}
\providecommand{\BIBentrySTDinterwordspacing}{\spaceskip=0pt\relax}
\providecommand{\BIBentryALTinterwordstretchfactor}{4}
\providecommand{\BIBentryALTinterwordspacing}{\spaceskip=\fontdimen2\font plus
\BIBentryALTinterwordstretchfactor\fontdimen3\font minus \fontdimen4\font\relax}
\providecommand{\BIBforeignlanguage}[2]{{%
\expandafter\ifx\csname l@#1\endcsname\relax
\typeout{** WARNING: IEEEtran.bst: No hyphenation pattern has been}%
\typeout{** loaded for the language `#1'. Using the pattern for}%
\typeout{** the default language instead.}%
\else
\language=\csname l@#1\endcsname
\fi
#2}}
\providecommand{\BIBdecl}{\relax}
\BIBdecl

\bibitem{o-ran-specification}
{O-RAN Working Group 2}, ``O-ran ai/ml workflow description and requirements–v1.01,'' O-RAN.WG2.AIML-v01.01 Technical Specification, April 2020.

\bibitem{9579445}
A.~Garcia-Saavedra and X.~Costa-Pérez, ``O-ran: Disrupting the virtualized ran ecosystem,'' \emph{IEEE Communications Standards Magazine}, vol.~5, no.~4, pp. 96--103, 2021.

\bibitem{9839628}
A.~S. Abdalla, P.~S. Upadhyaya, V.~K. Shah, and V.~Marojevic, ``Toward next generation open radio access networks: What o-ran can and cannot do!'' \emph{IEEE Network}, vol.~36, no.~6, pp. 206--213, 2022.

\bibitem{8466370}
F.~D. Calabrese, L.~Wang, E.~Ghadimi, G.~Peters, L.~Hanzo, and P.~Soldati, ``Learning radio resource management in rans: Framework, opportunities, and challenges,'' \emph{IEEE Communications Magazine}, vol.~56, no.~9, pp. 138--145, 2018.

\bibitem{9372298}
A.~Feriani and E.~Hossain, ``Single and multi-agent deep reinforcement learning for ai-enabled wireless networks: A tutorial,'' \emph{IEEE Communications Surveys Tutorials}, vol.~23, no.~2, pp. 1tr226--1252, 2021.

\bibitem{9812489}
P.~H. Masur, J.~H. Reed, and N.~K. Tripathi, ``Artificial intelligence in open-radio access network,'' \emph{IEEE Aerospace and Electronic Systems Magazine}, vol.~37, no.~9, pp. 6--15, 2022.

\bibitem{9430561}
L.~Maggi, A.~Valcarce, and J.~Hoydis, ``Bayesian optimization for radio resource management: Open loop power control,'' \emph{IEEE Journal on Selected Areas in Communications}, vol.~39, no.~7, pp. 1858--1871, 2021.

\bibitem{9903386}
A.~M. Nagib, H.~Abou-zeid, and H.~S. Hassanein, ``Toward safe and accelerated deep reinforcement learning for next-generation wireless networks,'' \emph{IEEE Network}, vol.~37, no.~2, pp. 182--189, 2023.

\bibitem{10071941}
L.~Bonati, M.~Polese, S.~D'Oro, S.~Basagni, and T.~Melodia, ``Intelligent closed-loop ran control with xapps in openran gym,'' in \emph{European Wireless 2022; 27th European Wireless Conference}, 2022, pp. 1--6.

\bibitem{9061001}
N.~Kato, B.~Mao, F.~Tang, Y.~Kawamoto, and J.~Liu, ``Ten challenges in advancing machine learning technologies toward 6g,'' \emph{IEEE Wireless Communications}, vol.~27, no.~3, pp. 96--103, 2020.

\bibitem{9931127}
P.~Li, J.~Thomas, X.~Wang, A.~Khalil, A.~Ahmad, R.~Inacio, S.~Kapoor, A.~Parekh, A.~Doufexi, A.~Shojaeifard, and R.~J. Piechocki, ``Rlops: Development life-cycle of reinforcement learning aided open ran,'' \emph{IEEE Access}, vol.~10, pp. 113\,808--113\,826, 2022.

\bibitem{9229155}
A.~T.~Z. Kasgari, W.~Saad, M.~Mozaffari, and H.~V. Poor, ``Experienced deep reinforcement learning with generative adversarial networks (gans) for model-free ultra reliable low latency communication,'' \emph{IEEE Transactions on Communications}, vol.~69, no.~2, pp. 884--899, 2021.

\bibitem{10.5555/2789272.2886795}
J.~Garc\'{\i}a and F.~Fern\'{a}ndez, ``A comprehensive survey on safe reinforcement learning,'' \emph{Journal of Machine Learning Research}, vol.~16, no.~1, p. 1437–1480, 2015.

\bibitem{9789336}
C.~T. Nguyen, N.~Van~Huynh, N.~H. Chu, Y.~M. Saputra, D.~T. Hoang, D.~N. Nguyen, Q.-V. Pham, D.~Niyato, E.~Dutkiewicz, and W.-J. Hwang, ``Transfer learning for wireless networks: A comprehensive survey,'' \emph{Proceedings of the IEEE}, vol. 110, no.~8, pp. 1073--1115, 2022.

\bibitem{9999297}
H.~Zhou, M.~Erol-Kantarci, and V.~Poor, ``Knowledge transfer and reuse: A case study of ai-enabled resource management in ran slicing,'' \emph{IEEE Wireless Communications}, pp. 1--10, 2022.

\bibitem{ferrus2020machine}
R.~A. Ferr{\'u}s~Ferr{\'e}, J.~P{\'e}rez~Romero, J.~O. Sallent~Roig, I.~Vil{\`a}~Mu{\~n}oz, and R.~Agust{\'\i}~Comes, ``Machine learning-assisted cross-slice radio resource optimization: Implementation framework and algorithmic solution,'' \emph{ITU journal on future and evolving technologies (ITU J-FET)}, vol.~1, no.~1, pp. 1--18, 2020.

\bibitem{9999295}
A.~Abouaomar, A.~Taik, A.~Filali, and S.~Cherkaoui, ``Federated deep reinforcement learning for open ran slicing in 6g networks,'' \emph{IEEE Communications Magazine}, vol.~61, no.~2, pp. 126--132, 2023.

\bibitem{10001658}
H.~Zhang, H.~Zhou, and M.~Erol-Kantarci, ``Federated deep reinforcement learning for resource allocation in o-ran slicing,'' in \emph{IEEE Global Communications Conference (GLOBECOM)}, 2022, pp. 958--963.

\bibitem{9771605}
N.~Hammami and K.~K. Nguyen, ``On-policy vs. off-policy deep reinforcement learning for resource allocation in open radio access network,'' in \emph{IEEE Wireless Communications and Networking Conference (WCNC)}, 2022, pp. 1461--1466.

\bibitem{9814869}
M.~Polese, L.~Bonati, S.~D'Oro, S.~Basagni, and T.~Melodia, ``Colo-ran: Developing machine learning-based xapps for open ran closed-loop control on programmable experimental platforms,'' \emph{IEEE Transactions on Mobile Computing}, vol.~22, no.~10, pp. 5787--5800, 2023.

\bibitem{10078092}
A.~Filali, B.~Nour, S.~Cherkaoui, and A.~Kobbane, ``Communication and computation o-ran resource slicing for urllc services using deep reinforcement learning,'' \emph{IEEE Communications Standards Magazine}, vol.~7, no.~1, pp. 66--73, 2023.

\bibitem{10008614}
F.~Lotfi, O.~Semiari, and F.~Afghah, ``Evolutionary deep reinforcement learning for dynamic slice management in o-ran,'' in \emph{IEEE Globecom Workshops (GC Wkshps)}, 2022, pp. 227--232.

\bibitem{10.1145/3551660.3560908}
J.~Saad, K.~Khawam, M.~Yassin, S.~Costanzo, and K.~Boulos, ``Crowding game and deep q-networks for dynamic ran slicing in 5g networks,'' in \emph{Proceedings of the 20th ACM International Symposium on Mobility Management and Wireless Access}, ser. MobiWac '22.\hskip 1em plus 0.5em minus 0.4em\relax New York, NY, USA: Association for Computing Machinery, 2022, p. 37–46.

\bibitem{9685808}
S.~Zhao, H.~Abou-zeid, R.~Atawia, Y.~S.~K. Manjunath, A.~B. Sediq, and X.-P. Zhang, ``Virtual reality gaming on the cloud: A reality check,'' in \emph{IEEE Global Communications Conference (GLOBECOM)}, 2021, pp. 1--6.

\bibitem{9627832}
L.~Bonati, S.~D'Oro, M.~Polese, S.~Basagni, and T.~Melodia, ``Intelligence and learning in o-ran for data-driven nextg cellular networks,'' \emph{IEEE Communications Magazine}, vol.~59, no.~10, pp. 21--27, 2021.

\bibitem{7891795}
O.~Sallent, J.~Perez-Romero, R.~Ferrus, and R.~Agusti, ``On radio access network slicing from a radio resource management perspective,'' \emph{IEEE Wireless Communications}, vol.~24, no.~5, pp. 166--174, 2017.

\bibitem{9524965}
A.~M. Nagib, H.~Abou-Zeid, and H.~S. Hassanein, ``Transfer learning-based accelerated deep reinforcement learning for 5g ran slicing,'' in \emph{IEEE 46th Conference on Local Computer Networks (LCN)}, 2021, pp. 249--256.

\bibitem{Leibovich}
T.~Leibovich-Raveh, D.~J. Lewis, S.~Al-Rubaiey~Kadhim, and D.~Ansari, ``A new method for calculating individual subitizing ranges,'' \emph{Journal of Numerical Cognition}, vol.~4, no.~2, pp. 429--447, Sep. 2018.

\bibitem{10075524}
A.~M. Nagib, H.~Abou-Zeid, and H.~S. Hassanein, ``Accelerating reinforcement learning via predictive policy transfer in 6g ran slicing,'' \emph{IEEE Transactions on Network and Service Management}, vol.~20, no.~2, pp. 1170--1183, 2023.

\bibitem{10172347}
Z.~Zhu, K.~Lin, A.~K. Jain, and J.~Zhou, ``Transfer learning in deep reinforcement learning: A survey,'' \emph{IEEE Transactions on Pattern Analysis and Machine Intelligence}, pp. 1--20, 2023.

\bibitem{8016642}
K.~A.M., F.~Hu, and S.~Kumar, ``Intelligent spectrum management based on transfer actor-critic learning for rateless transmissions in cognitive radio networks,'' \emph{IEEE Transactions on Mobile Computing}, vol.~17, no.~5, pp. 1204--1215, 2018.

\bibitem{10.5555/3294996.3295161}
A.~Barreto, W.~Dabney, R.~Munos, J.~J. Hunt, T.~Schaul, H.~van Hasselt, and D.~Silver, ``Successor features for transfer in reinforcement learning,'' in \emph{Proceedings of the 31st International Conference on Neural Information Processing Systems}, ser. NIPS'17.\hskip 1em plus 0.5em minus 0.4em\relax Red Hook, NY, USA: Curran Associates Inc., 2017, p. 4058–4068.

\bibitem{rusu2015policy}
A.~A. Rusu, S.~G. Colmenarejo, C.~Gulcehre, G.~Desjardins, J.~Kirkpatrick, R.~Pascanu, V.~Mnih, K.~Kavukcuoglu, and R.~Hadsell, ``Policy distillation,'' \emph{arXiv preprint arXiv:1511.06295}, 2015.

\bibitem{8540003}
R.~Li, Z.~Zhao, Q.~Sun, C.-L. I, C.~Yang, X.~Chen, M.~Zhao, and H.~Zhang, ``Deep reinforcement learning for resource management in network slicing,'' \emph{IEEE Access}, vol.~6, pp. 74\,429--74\,441, 2018.

\bibitem{sutton2018reinforcement}
R.~S. Sutton and A.~G. Barto, \emph{Reinforcement learning: An introduction}.\hskip 1em plus 0.5em minus 0.4em\relax MIT press, 2018.

\bibitem{schulman2017proximal}
J.~Schulman, F.~Wolski, P.~Dhariwal, A.~Radford, and O.~Klimov, ``Proximal policy optimization algorithms,'' \emph{arXiv preprint arXiv:1707.06347}, 2017.

\bibitem{nagib2023does}
A.~M. Nagib, H.~Abou-Zeid, and H.~S. Hassanein, ``How does forecasting affect the convergence of drl techniques in o-ran slicing?'' \emph{arXiv preprint arXiv:2309.00489}, 2023.

\end{thebibliography}

\section*{Biographies}
\begin{IEEEbiography}[{\includegraphics[width=1in,height=1.25in,clip,keepaspectratio]{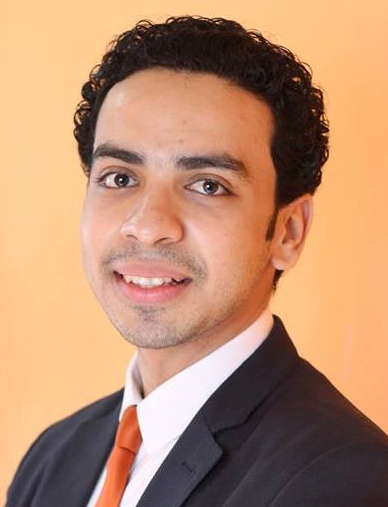}}]{Ahmad M. Nagib} (Graduate Student Member,
IEEE) received the B.Sc. and M.Sc. degrees from the Faculty of Computers and Artificial Intelligence, Cairo University. He is currently pursuing the Ph.D.
degree with the School of Computing, Queen’s University, where he is a Graduate Research Fellow at the Telecommunications Research Lab (TRL). He also works as an Assistant Lecturer with Cairo University and as a Machine Learning Ph.D. Co-op in the area of Cloud RAN with Ericsson, Canada. Mr. Nagib has recently been part of an industry-academia collaboration project with Ericsson, Canada. His research mainly addresses the practical challenges of applying machine learning, and specifically reinforcement learning, in next-generation wireless networks. Mr. Nagib's work has resulted in a record of publications in several IEEE flagship venues, such as the IEEE Journal on Selected Areas in Communications, IEEE Transactions on Network and Service Management, GLOBECOM, ICC, and LCN. He has also served as a Reviewer and a TPC Member in all of these respectable venues in addition to IEEE Transactions on Mobile Computing.
\end{IEEEbiography}

\begin{IEEEbiography}[{\includegraphics[width=1in,height=1.25in,clip,keepaspectratio]{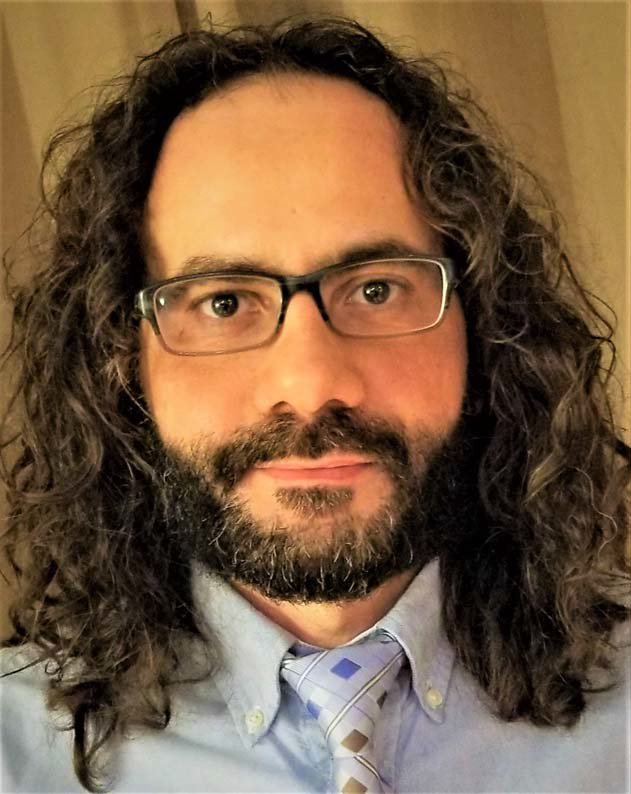}}]{Hatem Abou-Zeid} (Member, IEEE) is an Assistant Professor at the University of Calgary. Prior to that he was at Ericsson leading 5G radio access research and IP in RAN intelligence, low-latency communications, and spectrum sharing. Several wireless access and traffic engineering techniques that he co-invented and co-developed are deployed in 5G mobile networks worldwide. His research interests are broadly in 5G/6G networks, extended reality communications, and robust machine learning. His work resulted in 19 patent filings and over 60 journal and conference publications in several IEEE flagship venues.
He is an avid supporter of industry-university partnerships, and he served on the Ericsson Government Industry Relations and Talent Development Committees where he directed numerous academic research partnerships. He also served as Co-Chair of the IEEE ICC Workshop on Wireless Network Innovations for Mobile Edge Learning, and Corporate Co-Chair of the IEEE
LCN Conference 2022. Dr. Abou-Zeid received several awards for his academic contributions including a Best Paper Award at IEEE ICC 2022, the Software Engineering Professor of the Year and Early Research Excellence Awards in 2023 at the University of Calgary. He received the PhD degree from Queen’s University in 2014.
\end{IEEEbiography}

\begin{IEEEbiography}[{\includegraphics[width=1in,height=1.25in,clip,keepaspectratio]{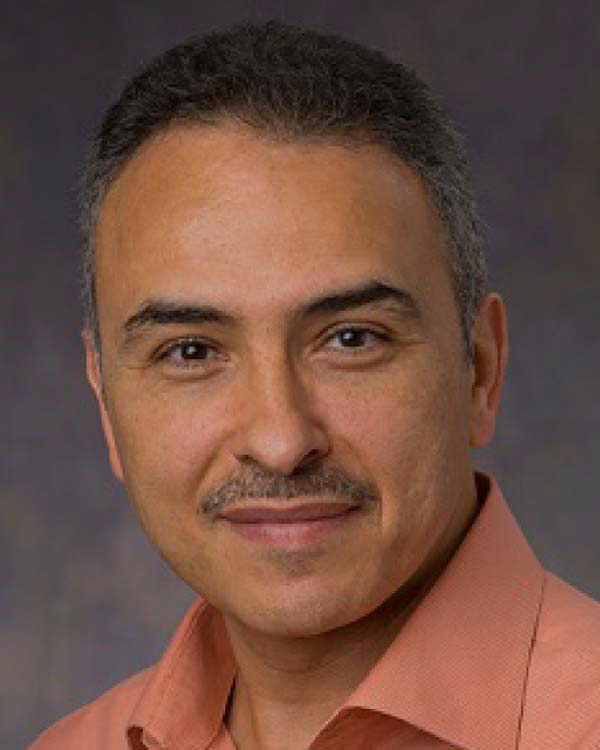}}]{Hossam S. Hassanein}
[S’86, M’90, SM’05, F’17] is a leading researcher in the areas of broadband, wireless and mobile networks architecture, protocols, control and performance evaluation. His record spans more than 600 publications in journals, conferences and book chapters, in addition to numerous keynotes and plenary talks in flagship venues. Dr. Hassanein has received several recognition and best paper awards at top international conferences. He is the founder and director of the Telecommunications Research Lab (TRL) at Queen's University School of Computing, with extensive international academic and industrial collaborations. He is the recipient of the 2016 IEEE Communications Society Communications Software Technical Achievement Award for outstanding contributions to routing and deployment planning algorithms in wireless sensor networks, and the 2020 IEEE IoT, Ad Hoc and Sensor Networks Technical Achievement and Recognition Award for significant contributions to technological advancement of the Internet of Things, ad hoc networks and sensing systems. Dr. Hassanein is a fellow of the IEEE, and is a former chair of the IEEE Communication Society Technical Committee on Ad hoc and Sensor Networks (TC AHSN). He is an IEEE Communications Society Distinguished Speaker (Distinguished Lecturer 2008-2010).
\end{IEEEbiography}

\end{document}